\documentclass[12pt]{article}
\usepackage{fullpage}
\usepackage{amsmath}
\usepackage{amsthm}
\usepackage{graphicx}
\usepackage[usenames]{color}
\newcommand{\bra}[1]{\langle #1|}
\newcommand{\ket}[1]{|#1\rangle}
\newcommand{\braket}[2]{\langle #1|#2\rangle}
\newcommand{\vnorm}[1]{\left|\left|#1\right|\right|}
\newcommand{\norm}[1]{\left|#1\right|}

\newtheorem{thm}{Theorem}[section]
\newtheorem{lem}[thm]{Lemma}

\title{The Adiabatic Theorem in the Presence of Noise}
\author{Michael J. O'Hara
\thanks{Graduate student in Applied Mathematics,
University of Maryland at College Park
(mjohara@gmail.com)}
\;and Dianne P. O'Leary
\thanks{Department of Computer Science and
Institute for Advanced Computer Studies, University
of Maryland; and
National Institute of Standards and Technology,
Mathematical and Computational Sciences Division,
Gaithersburg, Maryland.
Email: oleary@cs.umd.edu.
Research supported in part by National Science Foundation
Grant CCF 0514213.
}}

\begin{document}

\maketitle

\begin{abstract} 
  We provide rigorous bounds for the error of the adiabatic
  approximation of quantum mechanics under four sources of
  experimental error: perturbations in the initial condition,
  systematic time-dependent perturbations in the Hamiltonian, coupling
  to low-energy quantum systems, and decoherent time-dependent
  perturbations in the Hamiltonian.  For decoherent perturbations, we
  find both upper and lower bounds on the evolution time to guarantee
  the adiabatic approximation performs within a prescribed tolerance.
  Our new results include explicit definitions of constants, and we
  apply them to the spin-1/2 particle in a rotating magnetic field,
  and to the superconducting flux qubit.  We compare the theoretical
  bounds on the superconducting flux qubit to simulation results.
\end{abstract} 

\section{Introduction}

Adiabatic quantum computation \cite{farhi} (AQC) is a model of quantum
computation equivalent to the standard model \cite{aharonov}.  A
physical system is slowly evolved from the ground state of a simple
system, to one whose ground state encodes the solution to the
difficult problem.  By a physical principle known as the adiabatic
approximation, if the evolution is done sufficiently slowly, and the
minimum energy gap separating the ground state from higher states is
sufficiently large, then the final state of the system should be the
state encoding the solution to some problem.

The biggest hurdles facing many potential implementations of a quantum
computer are the errors due to interaction of the qubits with the
environment.  However, the effects of such errors are different in AQC
than in standard quantum computing. AQC is robust against dephasing in
the ground state, for instance \cite{childs}, and some have suggested
that noise in some regimes might actually assist adiabatic quantum
computation \cite{gaitan}.

The physical principle underlying AQC is established by the Adiabatic
Theorem (AT).  The AT bounds the run-time of the algorithm using the
minimum energy gap of the system during the evolution.  The AT itself
has been recently subject to controversy \cite{marzlin,tong,vertesi},
and cannot be applied directly to systems with noise or decoherence.
There have been some numerical studies of AQC in the presence of noise
\cite{childs, gaitan}, and an analytic random-matrix study
\cite{rolandnoise}.  Several recent studies have focused on the
adiabatic approximation in open quantum systems using the density
operator formalism \cite{sarandy,thunstrom,yi,fleischer,tiersch}.
However, it is difficult to derive rigorous bounds with this approach
because the dynamics involve a non-Hermitian operator without a
complete set of orthonormal eigenstates.

The subject of this work is the study of the adiabatic theorem in the
presence of noise, perturbations, and decoherence.  We consider
several ways to extend the application of the theorem.  Our
statements, derived from Avron's version of the theorem \cite{avron},
include explicit definitions of constants, so that we may apply them
on examples.

Experimental error for quantum computing experiments can be
conveniently divided into three categories \cite{noisetypes}:
\begin{enumerate}
\item {\em Coherent} errors, due to a systematic implementation error
  such as miscalibration in a magnetic field generator.
\item {\em Incoherent} errors, due to deterministic qubit-level
  differences in the evolution such as those caused by manufacturing
  defects.
\item {\em Decoherent} errors, which are are random qubit-level errors
  due to coupling with the environment.
\end{enumerate}
In Section~\ref{sec::noisyAT}, we prove several extensions of the
adiabatic theorem to handle these different types of error.  For
coherent errors, we provide a theorem for perturbations in the initial
state of the system, and a theorem for systematic time-dependent
perturbations in the Hamiltonian. In the case of decoherent errors, we
provide two new theorems, one for open quantum systems and one for
noise modeled as a time-dependent perturbation in the Hamiltonian.

In Section~\ref{sec::tongEX}, we apply the new theorems to the
spin-1/2 particle in a rotating magnetic field, a standard example for
controversy regarding the Adiabatic Theorem
\cite{bozic,tong,wuyang,mackenzieTong}.  We show our theorems make
correct predictions about the error of the adiabatic approximation.

Finally, in Section~\ref{sec::fluxqubit} we apply the new theorems to
the superconducting flux qubit \cite{orlando}, which has been proposed
for adiabatic quantum computation \cite{lloyd}.  We use our theorems
to determine a range of evolution times where the adiabatic
approximation is guaranteed to perform well for a typical set of
physical parameters and an apparently reasonable physical noise
source.  This provides the experimentalist with analytic tools for
determining parameters to guarantee the adiabatic approximation works
well, without the need to perform numerical simulations.

\section{Adiabatic Theorems for Noisy Hamiltonian Evolutions}
\label{sec::noisyAT}

We begin with a Hamiltonian evolution $\mathcal{H}(s)$ parameterized
by $s\in[0,1]$.  If we define $\tau$ to be the total evolution time,
then the Hamiltonian at time $t$ is $\mathcal{H}(t/\tau)$.  Thus, as
$\tau$ grows, $\mathcal{H}(s)$ describes a slower evolution.  Assume
$\mathcal{H}(s)$ has countable eigenstates $\{\ket{\psi_j(s)}\}$ and
eigenvalues $\lambda_0(s)\leq \lambda_1(s)...$, and consider the
subspace
\begin{equation}
  \Psi(s) = {\rm Span}\left\{\ket{\psi_m(s)},...,\ket{\psi_n(s)}\right\}\;,
\end{equation}
for some $0 \leq m \leq n$.  Then the {\em adiabatic approximation}
states that if the state of the system is contained in $\Psi(0)$ at
$t=0$, then at time $t=s/\tau$ the state is contained in $\Psi(s)$.
Notice that while the ground state $\ket{\psi_0(s)}$ may be important
for physical reasons, the definition above allows
consideration of a more general set of states.

It is convenient to define an operator that computes the error of the
adiabatic approximation for a Hamiltonian evolution.  We will need the
projection operator $P(s)$ that projects a state onto $\Psi(s)$.  We
will also need the unitary evolution operator $U(s)$, that is the
solution to Schr\"{o}dinger's equation in the form
\begin{equation}
  \dot{U}(s) = -i\tau \mathcal{H}(s) U(s) \;.
\end{equation}
To compute the error of the adiabatic approximation, we apply $P(0)$
to obtain the component of the initial state contained in $\Psi(0)$,
evolve it forward in time by applying $U(s)$, and then apply $I-P(s)$
to compute the component of the state outside $\Psi(s)$.  For
convenience, we define $Q(s)=I-P(s)$, so the error operator is
$Q(s)U(s)P(0)$.

In fact it will be most useful to bound the 2-norm of this operator,
denoted $\vnorm{Q(s)U(s)P(0)}$.  The 2-norm of an operator $A$ is the
square root of the largest eigenvalue of $A^{\dagger}A$, and in this
case yields a bound on the magnitude of the output state, given a
normalized input state.

The version of the adiabatic theorem that we use to bootstrap our
proof is based most closely on that of Reichardt \cite{reichardt},
which is based on that by Avron \cite{avron} (with later corrections
\cite{avron2} \cite{klein}).  The differences between our theorem and
Reichardt's theorem are
\begin{itemize}
  \item Our version of the theorem includes an explicit definition of
    constants, necessary to obtain quantitative bounds.
  \item Our version of the theorem applies to subspaces
    rather than only a non-degenerate state.
  \item We also present an integral formulation which provides better
    bounds when the energy gap is small for a very brief interval.
\end{itemize}
Throughout the paper, we use units where $\hbar=1$.

\begin{thm}[The Adiabatic Theorem (AT)]
Assume for $0\leq s\leq 1$ that $\mathcal{H}(s)$ is twice
differentiable, and let
\begin{align}
  \vnorm{\dot{\mathcal{H}}(s)} &\leq b_1(s) \;,&
  \vnorm{\ddot{\mathcal{H}}(s)} &\leq b_2(s) \;.
\end{align}
Further assume that $\mathcal{H}(s)$ has a countable number of
eigenstates, with eigenvalues \\$\lambda_0(s)\leq\lambda_1(s)...$, and
that $P(s)$ projects onto the eigenspace associated with the
eigenvalues $\{\lambda_m(s),\lambda_{m+1}(s),...\lambda_n(s)\}$.
Define
\begin{align}
  w(s) &= \lambda_n(s)-\lambda_m(s) \;,& 
  \gamma(s) &= \left\{ \begin{array}{ll}
      \min \{\lambda_{n+1}(s)-\lambda_n(s),
       \lambda_m(s)-\lambda_{m-1}(s)\} & m > 0\\
       \lambda_{n+1}(s)-\lambda_n(s) & m = 0
       \end{array} \right. \notag \\
  D(s) &= 1 + \frac{2w(s)}{\pi \gamma(s)} \;,&
  Q(s) &= I - P(s) \;.
\end{align}
Finally, assume $\gamma(s) >0 $ all $s$.  Then we have
\begin{align}
  \vnorm{Q(s)U(s)P(0)} 
 \leq& \frac{8D^2(0)b_1(0)}{\tau\gamma^2(0)}
  + \frac{8D^2(s)b_1(s)}{\tau\gamma^2(s)} \notag \\
  & + \int_{0}^s \frac{8D^2(r)}{\tau\gamma^2(r)}
  \left( \frac{8(1+D(r))b_1^2(r)}{\gamma(r)} + b_2(r)\right) dr \;.
  \label{eqn::AT}
\end{align}
\begin{proof}
  See Appendix A.
\end{proof}
\end{thm}

Notice that the first two terms in Equation (\ref{eqn::AT}) do not go
to zero as $s \rightarrow 0$, which is a consequence of
simplifications that were made to determine this bound.  However,
since AQC is the intended application of our results, we are only
interested in the error bound at the end of the evolution, namely
$s=1$.  Also, we will usually assume there are $\bar{b}_1\geq b_1(s)$,
$\bar{b}_2\geq b_2(s)$, $\bar{\gamma}\leq \gamma(s)$, and $\bar{D}
\geq D(s)$ for $s\in [0,1]$.  Then we can find a
constant upper bound for the integrand in Equation (\ref{eqn::AT}) and
thus bound the integral, resulting in the simpler
expression
\begin{align}
  \vnorm{Q(s)U(s)P(0)} 
   &\leq \frac{8\bar{D}^2}{\tau\bar{\gamma}^2} \left(
  2\bar{b}_1 +s\bar{b}_2
     + s\frac{8(1+\bar{D})\bar{b}_1^2}{\bar{\gamma}} \right)\;.
\end{align}
In fact, we will usually be interested in the AT for non-degenerate
ground states, in which case $m=n=0$ and $\bar{D}=1$, and we can use
the inequality
\begin{align}
  \vnorm{Q(s)U(s)P(0)} 
   &\leq \frac{8}{\tau\bar{\gamma}^2} \left(
      2\bar{b}_1 
      +s\bar{b}_2
      + s\frac{16\bar{b}_1^2}{\bar{\gamma}}
      \right)\;.
\end{align}

Also notice our statement of the AT is consistent with the common
interpretation of the theorem: if $\tau \gg 1/\bar{\gamma}^2$ then the
error in the adiabatic approximation is small.

\subsection{Coherent or Incoherent Errors}

Coherent or incoherent errors, due to systematic or deterministic
perturbations, may occur in one of two ways: either as a perturbation
in the initial condition or as a smooth perturbation in the
Hamiltonian. In this section, we see how such errors affect the
adiabatic approximation for a non-degenerate ground state.

Let us first consider a perturbation in the initial state,
\begin{equation}
  \ket{\phi(0)} = \eta \left(\ket{\psi_0(0)} + 
                             \delta \ket{\phi_{\perp}}\right) \;,
  \label{eqn::perturbinit}
\end{equation}
where $\eta^{-2} = 1+\norm{\delta}^2$ is a normalization factor,
$\ket{\psi_0(0)}$ is the ground state of $\mathcal{H}(0)$, and
$\ket{\phi_{\perp}}$ is some state orthogonal to $\psi_0(0)$.
It is not sufficient here to define the error of the
adiabatic approximation as the norm of the operator $Q(s)U(s)P(0)$,
where $P(s)$ is the projection onto $\ket{\psi_0(s)}$, since this does
not depend on the initial state.  The component of the final state
which lies outside the ground state at normalized time $s$ is
$Q(s)U(s)\ket{\phi(0)}$, and so here we take this as the
error.

\begin{thm}[AT for Error in the Initial State (AT-Initial)]
  Let $\mathcal{H}(s)$ have the properties required by the AT, and let
  the initial state $\ket{\phi(0)}$ be as in Equation
  (\ref{eqn::perturbinit}).  Then the error is bounded as
  \begin{equation}
    \vnorm{Q(s)U(s)\ket{\phi(0)}} \leq 
    \norm{\eta} \left( \norm{\delta} + 
    \frac{8}{\tau\bar{\gamma}^2} 
    \left(2\bar{b}_1+s\bar{b}_2
      +s\frac{16\bar{b}_1^2}{\bar{\gamma}}\right)\right) \;.
  \end{equation}
  \begin{proof}
    Using the AT and the triangle inequality for
    operator norms, and noting that the norm of unitary
    and projection operators is unity, we have
    \begin{align}
      \vnorm{Q(s)U(s)\ket{\phi(0)}} 
      &= \vnorm{ Q(s) U(s) \eta \left(\ket{\psi_0(0)} + 
	\delta \ket{\phi_{\perp}}\right)} \\
      &= \vnorm{ \eta \left(Q(s) U(s) P(0) \ket{\psi_0(0)} + 
        \delta Q(s) U(s) \ket{\phi_{\perp}}\right)} \\
      &\leq \norm{\eta} \left( \vnorm{ Q(s) U(s) P(0)} 
                              + \norm{\delta}\right) \\
      &\leq \norm{\eta} \left( \norm{\delta} 
	+ \frac{8}{\tau\bar{\gamma}^2} \left(2\bar{b}_1+s\bar{b}_2
	+s\frac{16\bar{b}_1^2}{\bar{\gamma}}\right)\right) \;.
    \end{align}
  \end{proof}
\end{thm}

Now suppose there is a smooth perturbation in the Hamiltonian caused
by a systematic error, so that
\begin{equation}
  \mathcal{H}_{\epsilon}(s) = \mathcal{H}(s) +\epsilon \Delta(s) \;.
\end{equation}
Then we can use the AT on
$\mathcal{H}_{\epsilon}$ by observing that
\begin{align}
  \vnorm{\dot{\mathcal{H}}_{\epsilon}(s)}
  &\leq \vnorm{\dot{\mathcal{H}}(s)} + 
      \epsilon \vnorm{\dot{\Delta}(s)} \;,\\
  \vnorm{\ddot{\mathcal{H}}_{\epsilon}(s)}
  &\leq \vnorm{\ddot{\mathcal{H}}(s)} + 
               \epsilon \vnorm{\ddot{\Delta}(s)}\;.
\end{align}
However, we must account for the difference in ground state between
$\mathcal{H}_{\epsilon}(s)$ and $\mathcal{H}(s)$.  Since we want to
measure error from the intended eigenstates of the system, not the
perturbed eigenstates, the error operator is $Q(s) U_{\epsilon}(s)
P(0)$, where we introduce the following notation:
\vskip 0.1in
\begin{tabular}{lp{5in}}
  $U_{\epsilon}(s)$ & The solution to
  $\dot{U}_{\epsilon}(s) = -i\tau
  \mathcal{H}_{\epsilon}(s)U_{\epsilon}(s)$.\\
  $P_{\epsilon}(s)$ & The projection operator onto the ground state
  of $\mathcal{H}_{\epsilon}(s)$. \\
  $Q_{\epsilon}(s)$ & $I - P_{\epsilon}(s)$. \\
  $\bar{\gamma}_{\epsilon}$ & The minimum energy gap between the
  ground state and first excited state of $\mathcal{H}_{\epsilon}(s)$.
\end{tabular}
\vskip 0.1in 

\begin{thm}[AT for Systematic Error (AT-Error)]
Assume that $\mathcal{H}_{\epsilon}(s)$ has the properties required by
the AT, and let
\begin{align}
  \vnorm{\dot{\mathcal{H}}_{\epsilon}(s)} &\leq \bar{b}_1 & 
  \vnorm{\ddot{\mathcal{H}}_{\epsilon}(s)} &\leq \bar{b}_2\\
  \sqrt{1-\norm{\braket{\psi_0(0)}{\phi_0(0)}}^2} &= \delta_0 &
  \sqrt{1-\norm{\braket{\psi_0(1)}{\phi_0(1)}}^2} &= \delta_1 \;,
\end{align}
where $\ket{\psi_0(s)}$ is the ground state of
$\mathcal{H}_{\epsilon}(s)$ and $\ket{\phi_0(s)}$ is the ground state
of $\mathcal{H}(s)$.  If $\bar{\gamma}_{\epsilon} > 0$, then we have
\begin{align}
  \vnorm{Q(1) U_{\epsilon}(1) P(0)} \leq 
  \frac{8}{\tau\bar{\gamma}_{\epsilon}^2}
  \left(2\bar{b}_1+\bar{b}_2+
  \frac{16\bar{b}_1^2}{\bar{\gamma}_{\epsilon}}\right) + 
  \delta_0 + \delta_1 + \delta_0 \delta_1 \; .
\end{align}
\begin{proof}

We know from AT that
\begin{equation}
  \vnorm{Q_{\epsilon}(1) U_{\epsilon}(1) P_{\epsilon}(0)} 
  \leq  \frac{8}{\tau\bar{\gamma}_{\epsilon}^2} 
  \left(2\bar{b}_1+\bar{b}_2+
    \frac{16\bar{b}_1^2}{\bar{\gamma}_{\epsilon}}\right) \;,
\end{equation}
but we want to find
$\vnorm{Q(1)U_{\epsilon}(1)P(0)}$. So define
\begin{equation}
  \Delta P(s) = P_{\epsilon}(s) - P(s) \;.
\end{equation}
Then we have
\begin{align}
  Q(1)U_{\epsilon}(1)P(0) 
  =& \left(Q_{\epsilon}(1)+\Delta P(1)\right) U_{\epsilon}(1) 
     \left(P_{\epsilon}(0)-\Delta P(0)\right) \\
  =& Q_{\epsilon}(1)U_{\epsilon}(1)P_{\epsilon}(0) 
     - Q_{\epsilon}(1)U_{\epsilon}(1)\Delta P(0)
     + \Delta P(1)U_{\epsilon}(1)P_{\epsilon}(0) \notag \\
   & - \Delta P(1)U_{\epsilon}(1)\Delta P(0) \;.
\end{align}
Now, the 2-norm of unitary and projection operators is unity, so
\begin{align}
  \vnorm{Q(1)U_{\epsilon}(1)P(0)} \leq \; &
  \frac{8}{\tau\bar{\gamma}_{\epsilon}^2}
    \left(2\bar{b}_1+\bar{b}_2+
    \frac{16\bar{b}_1^2}{\bar{\gamma}_{\epsilon}}\right)
    \notag \\
  & + \vnorm{\Delta P(0)} + \vnorm{\Delta P(1)} + 
    \vnorm{\Delta P(0)}\vnorm{\Delta P(1)} \; .  \label{eqn::intermed}
\end{align}
It remains to find $\vnorm{\Delta P(s)}$.  We hope to write
\begin{equation}
  \ket{\phi_0(s)} = M(s) \ket{\psi_0(s)}  \label{eqn::M}
\end{equation}
for some unitary transformation $M(s)$ that is close to the identity
provided $\psi_0(s)$ and $\phi_0(s)$ are close to each other.  We use
the Givens rotation, where the first basis state is $\ket{\phi_0}$ and
the second is the complement of the projection of $\ket{\psi_0}$ onto
the first basis state:
\begin{align}
  \hat{e}_1 &= \ket{\phi_0(s)} & \hat{e}_2 &= \frac{
  \left(1-\ket{\phi_0(s)}\bra{\phi_0(s)}\;\right) \;\ket{\psi_0(s)}}
  {\sqrt{1-\norm{\braket{\phi_0(s)}{\psi_0(s)}}^2}} \;.
\end{align}
The remaining basis states are chosen arbitrarily so long as the
resulting basis is orthonormal and spans the Hilbert space.  In that
basis, Equation (\ref{eqn::M}) is realized by
\begin{align}
  \left( \begin{array}{lrcccc}
     C^*(s) & S(s)  &   &   &        & \\
     -S(s)  & C (s) &   &   &        & \\
            &       & 1 &   &        & \\
            &       &   & 1 &        & \\
            &       &   &   & \ddots & \\
            &       &   &   &        & 1
  \end{array} \right) 
  \left( \begin{array}{c}
    C(s) \\
    S(s) \\
    0 \\
    0 \\
    \vdots \\
    0
  \end{array} \right) 
  =
  \left( \begin{array}{c} 
    1 \\
    0 \\
    0 \\
    0 \\
    \vdots \\
    0
 \end{array} \right) \;,
\end{align}
where we define
\begin{align}
  C(s) &= \braket{\phi_0(s)}{\psi_0(s)} \\
  S(s) &= \sqrt{1-\norm{\braket{\phi_0(s)}{\psi_0(s)}}^2} \;.
\end{align}
We see that
\begin{align}
  M(s) P_{\epsilon}(s) M^{\dagger}(s)
  &= M(s) \ket{\psi_0(s)} \bra{\psi_0(s)} M^{\dagger}(s) \\
  &= \ket{\phi_0(s)} \bra{\phi_0(s)}\\
  &= P(s) \;.
\end{align}
Letting $E(s)=I-M(s)$, we have
\begin{align}
  \Delta P(s) &= P_{\epsilon}(s) - P(s) \\
  &=  M^{\dagger}(s) P(s) M(s) -
                         M^{\dagger}(s) M(s) P(s) \\
	&= M^{\dagger}(s) \left[ P(s), M(s) \right] \\
        &= M^{\dagger}(s) \left[ E(s), P(s) \right] \;.
\end{align} 
But we know $E(s)$ and $P(s)$:
\begin{align}
  E(s) &= 
  \left( \begin{array}{lrcccc}
     1-C^*(s) & -S(s)  &   &   &        & \\
     S(s)     & 1-C(s) &   &   &        & \\
              &        & 0 &   &        & \\
              &        &   & 0 &        & \\
              &        &   &   & \ddots & \\
              &        &   &   &        & 0
  \end{array} \right) & 
  P(s) &= \left( \begin{array}{lrcccc} 
    1 & 0 &   &   &        & \\ 
    0 & 0 &   &   &        & \\
      &   & 0 &   &        & \\
      &   &   & 0 &        & \\
      &   &   &   & \ddots & \\
      &   &   &   &        & 0
  \end{array} \right) \;,
\end{align} 
so 
\begin{align}
  [E(s),P(s)] = 
  \left( \begin{array}{lrcccc}
    0 & S(s) &   &   &        & \\
    S(s) & 0 &   &   &        & \\
      &   & 0 &   &        & \\
      &   &   & 0 &        & \\
      &   &   &   & \ddots & \\
      &   &   &   &        & 0   
  \end{array} \right) \;.
\end{align}
Finally,
\begin{align}
  \vnorm{\Delta P(s)} &= \vnorm{P_{\epsilon}(s) - P(s)} \\
  &= \vnorm{\;\left[ E(s), P(s) \right]\;} \\
  &= S(s) \;. \label{eqn::DPNorm}
\end{align}
Combining Equations (\ref{eqn::DPNorm}) and (\ref{eqn::intermed})
yields the theorem.
\end{proof}
\end{thm}

When it is inconvenient to compute $\delta_0$ and $\delta_1$ exactly,
they can be bounded using the ``$\sin(\Theta)$
theorem'' \cite[page 251]{stewart}:
\begin{align}
  \delta_0 & \leq \frac{\epsilon\vnorm{ \Delta(0)}}
	{\lambda_1(0) - \lambda_0^{\epsilon}(0)} &
	\delta_1 & \leq \frac{\epsilon\vnorm{ \Delta(1)}}
	      {\lambda_1(1) - \lambda_0^{\epsilon}(1)}  \;,
\end{align}
where $\lambda_0^{\epsilon}(s)$ is the energy of the ground state of
$\mathcal{H}_{\epsilon}(s)$.  If $\lambda_0^{\epsilon}(s)$ is
difficult to find, we can use the Bauer-Fike theorem
\cite[page 192]{stewart} to get
\begin{align}
  \delta_0 & \leq \frac{\epsilon\vnorm{ \Delta(0)}}
	{\gamma(0) - \epsilon\vnorm{\Delta(0)}} &
	\delta_1 & \leq \frac{\epsilon\vnorm{ \Delta(1)}}
	      {\gamma(1)- \epsilon\vnorm{\Delta(1)}}\;,
\end{align}
where $\gamma(s)$ is the energy gap of the unperturbed Hamiltonian
$\mathcal{H}(s)$.

A remarkable feature of AT-Error is that it does
not depend directly on the magnitude of the perturbation term
$\epsilon\Delta(s)$ except at the endpoints.  It does not matter which
path we take through state space, so long as we begin and end near the
correct Hamiltonians and do not accumulate too much error along the
way.

\subsection{Decoherent Errors}

Now we consider decoherent errors induced, perhaps, by noise in the
environment.  We first consider noise modeled as a coupled quantum
system where the environment Hamiltonian is
independent of time, and then as a classical
time-dependent perturbation in the
Hamiltonian.

For the environment Hamiltonian $\mathcal{H}_{env}$ and interaction
Hamiltonian $\epsilon \Delta(s)$, we can write the combined
Hamiltonian $\mathcal{H}_{\epsilon}(s)$ as
\begin{equation}
  \mathcal{H}_{\epsilon}(s) = \mathcal{H}(s) \otimes I 
                              + I \otimes \mathcal{H}_{env}
			      + \epsilon \Delta(s)\;.
\end{equation}
Direct application of the AT yields a very pessimistic result because
the ground state of the composite system has, in the weak coupling
limit, both the target system and the environment in the ground state
of their respective Hamiltonians.  The target system remaining in the
ground state and the environment tunneling to its first excited state
will be considered a failure of the adiabatic approximation.  An
experimentalist probably cannot achieve the environmental ground
state, and the energy gap between environment states is likely quite
small so that the AT produces a very large error bound.

One way to resolve this problem in the interpretation of the adiabatic
approximation is to work with the density operator that results from
the partial trace\cite{sarandy}:
\begin{equation}
  \rho(s) = {\rm Tr}_{env} (\rho_{\epsilon}(s)) \;,
\end{equation}
where $\rho_{\epsilon}(s)$ is the density operator associated with the
state of the composite system $\mathcal{H}_{\epsilon}(s)$.
Usually we can write
\begin{equation}
  \dot{\rho}(s) = \mathcal{L}(s)\rho(s)\;,
\end{equation}
where $\mathcal{L}(s)$ is a linear operator but not generally
Hermitian.  We might try to use this differential equation to prove an
adiabatic theorem restated in terms of the expectation
$\bra{\phi_0(s)}\rho(s)\ket{\phi_0(s)}$, where $\ket{\phi_0(s)}$ is
the ground state of $\mathcal{H}(s)$.  The problem is that
$\mathcal{L}(s)$ does not have a complete set of orthonormal
eigenstates, which is of great assistance in proving the AT.  A
rigorous bound on the error of the adiabatic approximation has yet to
be found using this approach \cite{sarandy,yi,fleischer,thunstrom,tiersch}.

The density operator approach sums together the set of states in the
composite system whose measurement on the system of interest yields
the ground state.  Instead, below we will simply consider that set of
states a subspace, and identify conditions where the usual AT for
evolution of a subspace applies.

For adiabatic quantum computation, we expect the energy gap
$\bar{\gamma}$ to be significantly larger than the temperature $k_{B}
T$.  When the system is significantly coupled to only a small number
$N$ of nearby particles, then the range of relevant environmental
energy levels is on the order of $N k_{B} T$, so
$\vnorm{\mathcal{H}_{env}}$ is also order $Nk_BT$. If
$\vnorm{\mathcal{H}_{env}} < \bar{\gamma}$, we may use the AT.

More generally, we must determine the error operator for evolution in
the composite system.  The projection operator of interest projects
states in the composite system onto those states whose measurement
reveals the original system to be in the ground state.  If $P(s)$ is
the projection onto the ground state of $\mathcal{H}(s)$, then this
operator is $P(s)\otimes I$.  Its complement is $ Q(s) \otimes I =
I\otimes I - P(s)\otimes I$, since Kronecker products distribute.
Then the error operator is $(Q(s)\otimes I) \;U_{\epsilon}(s)\;
(P(0)\otimes I)$.

\begin{thm}[AT for Coupling to Low-Temperature Environment (AT-Env)]
Suppose we are given
\begin{equation}
  \mathcal{H}_{\epsilon}(s) = \mathcal{H}(s)\otimes I 
         + I\otimes \mathcal{H}_{env} +\epsilon\Delta(s) \;,
\end{equation}
and suppose we can choose $w$ so that
\begin{equation}
  \vnorm{\mathcal{H}_{env}} + 2\epsilon\vnorm{ \Delta(s)}
  \leq w < \bar{\gamma} \;,
\end{equation}
where $\bar{\gamma}$ is the minimum energy gap between the ground
state and first excited state of $\mathcal{H}(s)$.  Assume that
$\mathcal{H}_{\epsilon}(s)$ has the properties required by the AT, and
assume that $\mathcal{H}_{env}$ has $M$ states and its
ground state has zero energy.  Let
\begin{align}
  \vnorm{\dot{\mathcal{H}}_{\epsilon}(s)} &\leq \bar{b}_1 \;,& 
  \vnorm{\ddot{\mathcal{H}}_{\epsilon}(s)} &\leq \bar{b}_2 \;,\\
  \delta_0 &= \frac{\epsilon\vnorm{ \Delta(0)}}
       {\bar{\gamma} - \vnorm{\mathcal{H}_{env}} 
	 - \epsilon\vnorm{ \Delta(0)}} \;, &
  \delta_1 &= \frac{\epsilon\vnorm{ \Delta(1)}}
       {\bar{\gamma} - \vnorm{\mathcal{H}_{env}} 
       - \epsilon\vnorm{ \Delta(1)}} \;, \\
  \bar{\gamma}_{\epsilon} &= 
  \left\{ \begin{array}{ll}
           \bar{\gamma}-w &: \epsilon > 0 \\
	   \bar{\gamma}   &: \epsilon = 0
  \end{array} \right. \;, &
       \bar{D} &= 
       \left\{ \begin{array}{ll}
	 1 + \frac{2w}{\pi \bar{\gamma}_{\epsilon}} & : \epsilon > 0 \\
	 1 & : \epsilon = 0
       \end{array}\right.
       \;.
\end{align}
Then we have
\begin{align}
  \vnorm{(Q(1)\otimes I) \;U_{\epsilon}(1)\;(P(0)\otimes I)} 
    &\leq \frac{8\bar{D}^2}{\tau\bar{\gamma}_{\epsilon}^2} \left(
      2\bar{b}_1 
      +\bar{b}_2
      + \frac{8(1+\bar{D})\bar{b}_1^2}{\bar{\gamma}_{\epsilon}}
      \right)
      +\delta_0 + \delta_1 + \delta_0 \delta_1 \;,
\end{align}
where $\tau$ is the total evolution time.
\begin{proof}

For $\epsilon = 0$, we can ignore $\mathcal{H}_{env}$ and this 
theorem is simply the AT.  So let us consider $\epsilon > 0$.  We will
do this by considering $\epsilon>0$ as a perturbation of the $\epsilon
=0$ case.

For $\epsilon = 0$, the eigenstates of $\mathcal{H}_{\epsilon}(s)$ are
simply the eigenstates of $\mathcal{H}(s)$ tensored to the eigenstates
of $\mathcal{H}_{env}$, and the energy of those states is the sum of
the energy of the state in $\mathcal{H}(s)$ and the energy of the
state in $\mathcal{H}_{env}$.

Define the ground state energy of $\mathcal{H}(s)$ as $\lambda_0(s)$,
the energy of the first excited state as $\lambda_1(s)$, and
$\gamma(s) = \lambda_1(s)-\lambda_0(s)$.  Recall that
the $M$ energies of the $\mathcal{H}_{env}$ states are
non-negative and less than $\bar{\gamma}$.  Then the first $M$
eigenstates of $\mathcal{H}_{\epsilon}(s)$ are the ground state of
$\mathcal{H}(s)$ tensored with different eigenstates of
$\mathcal{H}_{env}$, and the rest of the states are some excited state
of $\mathcal{H}(s)$ tensored with an environment state.

In particular, the $M^{th}$ state of $\mathcal{H}_{\epsilon}$ is the
ground state of $\mathcal{H}$ tensored with the most energetic state
of $\mathcal{H}_{env}$, and thus has energy $\lambda_0(s) +
\vnorm{\mathcal{H}_{env}}$.  The $M+1$ state is the first excited
state of $\mathcal{H}$ tensored with the ground state of
$\mathcal{H}_{env}$, and has energy $\lambda_1(s)$.  So the energy gap
between the first $M$ states and the rest of the spectrum is is
exactly $\gamma(s)-\vnorm{\mathcal{H}_{env}}$.

For positive $\epsilon$, these eigenstates are perturbed.  Using the
Bauer-Fike theorem \cite[page
192]{stewart}, we see that the gap is reduced by at most
$2\epsilon\vnorm{\Delta(s)}$ in the
presence of coupling, so the gap is still at least
$\bar{\gamma}_{\epsilon}$.

What we want is the adiabatic approximation of the evolution of the
subspace formed by these $M$ eigenstates, with a spectral width at
most $w$ and an energy gap of $\bar{\gamma}_{\epsilon}$.  Following
the proof of AT-Error, define
\begin{equation}
  \Delta P(s) = P_{\epsilon}(s) - P(s)\otimes I\;,
\end{equation}
then we have
\begin{align}
  (Q(1)\otimes I)\;U_{\epsilon}(1)\;(P(0)\otimes I)
  =& \left(Q_{\epsilon}(1)
    +\Delta P(1)\right) U_{\epsilon}(1) 
     \left(P_{\epsilon}(0)-\Delta P(0)\right) \\
  =& Q_{\epsilon}(1)U_{\epsilon}(1)P_{\epsilon}(0) 
     - Q_{\epsilon}(1)U_{\epsilon}(1)\Delta P(0)
     + \Delta P(1)U_{\epsilon}(1)P_{\epsilon}(0) \notag \\
     & - \Delta P(1)U_{\epsilon}(1)\Delta P(0) \;,
\end{align}

We can bound $\vnorm{\Delta P(s)}$ using the fact that the singular
values of $\Delta P(s)$ are given by the sines of the canonical angles
between $P_{\epsilon}(s)$ and $P(s)\otimes I$
\cite[page 43]{stewart}, the
``$\sin(\Theta)$ theorem'' \cite[page
251]{stewart}, and the Bauer-Fike theorem
\cite[page 192]{stewart}:
\begin{equation}
  \Delta P(s) \leq \frac{\epsilon\vnorm{ \Delta(s)}}
	 {\gamma(s) - \vnorm{\mathcal{H}_{env}} 
	   - \epsilon\vnorm{ \Delta(s)}} \;,
\end{equation}
so 
\begin{align}
  \Delta P(0) &\leq \delta_0 & \Delta P(1) &\leq \delta_1 \;.
\end{align}
Now we are ready to apply the AT:
\begin{align}
  \vnorm{(Q(1)\otimes I)\;U_{\epsilon}(1)\;(P(0)\otimes I)}
   &\leq \vnorm{Q_{\epsilon}(1)U_{\epsilon}(1)P_{\epsilon}(0)}
   + \delta_0 + \delta_1 + \delta_0\delta_1\\
   &\leq  \frac{8\bar{D}^2}{\tau\bar{\gamma}_{\epsilon}^2} \left(
      2\bar{b}_1 
      +\bar{b}_2
      + \frac{8(1+\bar{D})\bar{b}_1^2}{\bar{\gamma}_{\epsilon}}
      \right)
 + \delta_0 + \delta_1 + \delta_0\delta_1\;.
\end{align}
\end{proof}
\end{thm}

Now let us consider another model of decoherent noise, namely a
time-dependent perturbation in the Hamiltonian.  There is a problem
applying the AT directly, because the time-dependent
perturbation is a function of true time $t$, not the
unitless evolution parameter $s$.  So as $\tau$ grows, more noise
fluctuations are packed into the interval $s\in[0,1]$, causing
$\vnorm{d\mathcal{H}/ds}$ to diverge.  Then there is no bound
$\bar{b}_1$ greater than $\vnorm{d\mathcal{H}/ds}$ that is independent
of $\tau$.  In fact this problem was the source of confusion in the
recent controversy surrounding the adiabatic theorem
\cite{marzlin,tong,vertesi}.

We will need to consider Hamiltonians $\mathcal{H}_{\tau}(s)$ that
depend on both $s$ and $t$.  We define the following notation:

\vskip 0.1in
\begin{tabular}{ll}
  $U_{\tau}(s)$ & The solution to $\dot{U}_{\tau}(s) =
  -i\tau \mathcal{H}_{\tau}(s)U_{\tau}(s)$ for a fixed
  $\tau$.\\
  $P_{\tau}(s)$ & The projection operator onto the ground state of
  $\mathcal{H}_{\tau}(s)$. \\ 
  $Q_{\tau}(s)$ & $I - P_{\tau}(s)$. \\
  $\gamma_{\tau}(s)$ & The energy difference between the ground state 
                 and first excited state of $\mathcal{H}_{\tau}(s)$.
\end{tabular}
\vskip 0.1in 

\begin{thm}[Adiabatic Theorem for Hamiltonian Evolutions on
            Two Time Scales (AT-2)] 

Suppose, for any fixed $\tau$, that $\mathcal{H}_{\tau}(s)$ has the
properties required by the AT.  Further assume there are real
functions $g_1(\tau)$ and $g_2(\tau)$ such that
\begin{align}
  \vnorm{\dot{\mathcal{H}}_{\tau}(s)} &\leq g_1(\tau) &
  \vnorm{\ddot{\mathcal{H}}_{\tau}(s)} &\leq g_2(\tau) \;,
\end{align}
for all $\tau$.  If there is a
$\bar{\gamma}_{min}$ so that
\begin{equation}
  0 < \bar{\gamma}_{min} \leq \gamma_{\tau}(s)
\end{equation}
for all $s$ and $\tau$, then we have
\begin{equation}
  \vnorm{Q_{\tau}(s)U_{\tau}(s)P_{\tau}(0)} \leq 
    \frac{8}{\tau \bar{\gamma}_{min}^2}
    \left(2g_1(\tau)+sg_2(\tau)+ 
    s\frac{16g_1^2(\tau)}{\bar{\gamma}_{min}}\right) \;.
\end{equation}
\begin{proof}
The lemma we are trying to prove is the union of special cases of
the AT, when the AT is applied to one-parameter projections of the
original Hamiltonian.

For fixed $\tau$, we consider $\mathcal{H}_{\tau}(s)$ as a
one-parameter Hamiltonian to which the usual AT will apply.  Then by
the AT, we can write
\begin{equation}
  \vnorm{Q_{\tau}(s) U_{\tau}(s) P_{\tau}(0)} \leq 
       \frac{8}{{\tau} \bar{\gamma}_{min}^2}
       \left( 2g_1({\tau})+sg_2({\tau})+
              s\frac{16g_1^2({\tau})}{\bar{\gamma}_{min}} \right) \;.
\label{eq::special}
\end{equation}
But we can do this for any $\tau$, so the lemma holds.
\end{proof}
\end{thm}

Now we can apply AT-2 to the case where there is an evolution
performed on some scaled time $s$, with an additive noise Hamiltonian
$\mathcal{H}_{noise}(t)$ that is a function of real time $t=s\tau$:
\begin{equation}
  \mathcal{H}_{\tau}(s) = \mathcal{H}(s) + \mathcal{H}_{noise}(s\tau) \;.
\end{equation}
We define the error operator for the noisy Hamiltonian as
$Q(s)U_{\tau}(s)P(0)$.  The projection operators refer to the
unperturbed Hamiltonian because success should be defined in terms of
the intended states.

\begin{thm}[Adiabatic Theorem for Noisy Hamiltonian
            Evolutions (AT-Noise)]

Suppose for any fixed $\tau$, that $\mathcal{H}_{\tau}(s) =
\mathcal{H}(s)+\mathcal{H}_{noise}(s\tau)$ has the properties required
by the AT.  Assume
\begin{align}
  \vnorm{\frac{d}{ds} \mathcal{H}(s)} &\leq c_1 & 
  \vnorm{\frac{d^2}{ds^2} \mathcal{H}(s)} &\leq c_2\\
  \vnorm{\frac{d}{dt} \mathcal{H}_{noise}(t)} &\leq d_1 & 
  \vnorm{\frac{d^2}{dt^2} \mathcal{H}_{noise}(t)} &\leq d_2 \\
  \sqrt{1-\norm{\braket{\psi_0(0)}{\phi_0(0)}}^2} &= \delta_0 &
  \sqrt{1-\norm{\braket{\psi_0(1)}{\phi_0(1)}}^2} &= \delta_1 \;,
\end{align}
where $\ket{\psi_0(s)}$ is the ground state of $\mathcal{H}_{\tau}(s)$
and $\ket{\phi_0(s)}$ is the ground state of $\mathcal{H}(s)$.
Further assume that there is a $\bar{\gamma}_{noise}$
so that
\begin{equation}
  0 < \bar{\gamma}_{noise} \leq \gamma_{\tau}(s)
\end{equation}
for all $s$ and $\tau$.  Then we have
\begin{align}
  \vnorm{Q(1)U_{\tau}(1)P(0)} \leq \; &
  \frac{8}{\bar{\gamma}_{noise}^2} \left[
    \left(d_2 + \frac{16d_1^2}{\bar{\gamma}_{noise}}\right) \tau +
    2d_1\left(1+\frac{16c_1}{\bar{\gamma}_{noise}}\right) + 
    \left(2c_1+c_2+\frac{16c_1^2}{\bar{\gamma}_{noise}}\right)\frac{1}{\tau}
  \right] \notag \\
  & + \delta_0 + \delta_1 + \delta_0 \delta_1 \; .
\end{align}

\begin{proof}
Evidently
\begin{align}
  \frac{d}{d s} \mathcal{H}_{\tau}(s) &= 
     \frac{d}{d s} \mathcal{H}(s) + 
     \tau \frac{d}{d t}\mathcal{H}_{noise}(t) \;,\\
  \vnorm{\frac{d}{d s} \mathcal{H}_{\tau}(s)}  &\leq c_1+\tau d_1 \;,
     \label{eqn::d1norm}\\
  \frac{d^2}{d s^2} \mathcal{H}_{\tau}(s) &= 
    \frac{d^2}{d s^2} \mathcal{H}(s) + 
    \tau^2 \frac{d^2}{d t^2}\mathcal{H}_{noise}(t) \;,\\
  \vnorm{\frac{d^2}{d s^2} \mathcal{H}_{\tau}(s)}  &\leq c_2+\tau^2 d_2 \;.
  \label{eqn::d2norm}
\end{align}
Substitution of Equation~(\ref{eqn::d1norm}) and
Equation~(\ref{eqn::d2norm}) into AT-2 yields, for all $s$ and $\tau$,
\begin{align}
  \vnorm{Q_{\tau}(s)U_{\tau}(s)P_{\tau}(0)} \leq \; &
  \frac{8}{\bar{\gamma}_{noise}^2} \left[
    \left(d_2 + \frac{16d_1^2}{\bar{\gamma}_{noise}}\right) \tau +
    2d_1\left(1+\frac{16c_1}{\bar{\gamma}_{noise}}\right) + 
    \left(2c_1+c_2+\frac{16c_1^2}{\bar{\gamma}_{noise}}\right)\frac{1}{\tau}
  \right]  \; .
\end{align}
As in the proof of AT-Error, we define
\begin{align}
  \Delta P(0) &= P_{\tau}(0) - P(0) & 
  \Delta P(1) &= P_{\tau}(1) - P(1) \;.
\end{align}
Then for $s=1$ we have
\begin{align}
  Q(1)U_{\tau}(1)P(0) 
  =& \left(Q_{\tau}(1)+\Delta P(1)\right) U_{\tau}(1) \left(P_{\tau}(0)-\Delta P(0)\right) \\
  =& Q_{\tau}(1)U_{\tau}(1)P_{\tau}(0) - Q_{\tau}(1)U_{\tau}(1)\Delta P(0)
  + \Delta P(1)U_{\tau}(1)P_{\tau}(0) \notag \\
  & - \Delta P(1)U_{\tau}(1)\Delta P(0) \;.
\end{align}
 Now we bound the norm of the error:
\begin{align}
  \vnorm{Q(1)U_{\tau}(1)P(0)} \leq \; &
  \frac{8}{\bar{\gamma}_{noise}^2} \left[
    \left(d_2 + \frac{16d_1^2}{\bar{\gamma}_{noise}}\right) \tau +
    2d_1\left(1+\frac{16c_1}{\bar{\gamma}_{noise}}\right) + 
    \left(2c_1+c_2+\frac{16c_1^2}{\bar{\gamma}_{noise}}\right)\frac{1}{\tau}
  \right] \notag \\
  & + \vnorm{\Delta P(0)} + \vnorm{\Delta P(1)} + 
  \vnorm{\Delta P(0)}\vnorm{\Delta P(1)} \; . \label{eqn::intermed2}
\end{align}
Using the Givens rotation just as in the proof of AT-Error, we have
\begin{align}
  \vnorm{\Delta P(0)} &= \delta_0 &
  \vnorm{\Delta P(1)} &= \delta_1\;,
\end{align}
which, when substituted into Equation (\ref{eqn::intermed2}),
completes the proof.
\end{proof}
\end{thm}

Several observations can be made about this result:
\begin{enumerate}
\item As with AT-Error, if it is inconvenient to compute $\delta_0$
  and $\delta_1$ exactly, they can be bounded using the
  ``$\sin(\Theta)$ theorem'' combined with the Bauer-Fike theorem
  \cite[page 192]{stewart}:
  \begin{align}
    \delta_0 & \leq \frac{\vnorm{\mathcal{H}_{noise}(0)}}
	  {\gamma(0) - \vnorm{\mathcal{H}_{noise}(0)}} &
	  \delta_1 & \leq \frac{\vnorm{\mathcal{H}_{noise}(1)}}
		{\gamma(1) - \vnorm{\mathcal{H}_{noise}(1)}} \;,
  \end{align}
  where $\gamma(s)$ is the energy gap between the ground state and
  first excited state of $\mathcal{H}(s)$.  Also, $\delta_0$ and
  $\delta_1$ can be taken as zero if $\mathcal{H}_{noise}(0) =
  \mathcal{H}_{noise}(\tau) = 0$.  In general, we expect them to be
  quite small if $\mathcal{H}_{noise}(t)$ is several orders of
  magnitude smaller than $\mathcal{H}(s)$.
\item When $\tau$ is small, the $1/\tau$ term dominates.  This term is
  exactly the bound from the (noiseless) AT.  It shows there is always
  a positive lower bound on the running time of the adiabatic
  algorithm to guarantee a particular error tolerance.
\item When $\tau$ is large, the first term dominates.  In fact, we can
  see that in the presence of noise, there is {\em always} some
  sufficiently large $\tau$ beyond which the adiabatic approximation
  may perform poorly.  So given an error tolerance, there is always an
  upper bound on the running time for the adiabatic algorithm, beyond
  which the theorem cannot guarantee the tolerance to be met.
\item If there is a great deal of noise, and thus $d_1$ is large, the
  constant term (with respect to $\tau$) could become as large as
  $\mathcal{O}(1)$ and there could be {\em no} running time for the
  adiabatic algorithm which results in an accurate calculation.
\end{enumerate}

We are also interested in a lower bound on the error of the adiabatic
approximation in the presence of noise.  A lower bound could be used
to prove that a certain amount of noise was unacceptable for AQC,
because it would guarantee failure of the algorithm for some level of
noise.  However, it will be difficult to get a non-trivial lower
bound, since there are time-dependent perturbations which yield {\em
zero} error in the adiabatic approximation, better than might exist
without the perturbation.  To see this, define
\begin{equation}
  \mathcal{H}_A(s) = \mathcal{H}(s)+i/\tau[\dot{P}(s),P(s)] \;,
\end{equation}
where the term $ i/\tau[\dot{P}(s),P(s)]$ is the perturbation. Avron
proved \cite{avron}, as do we in the appendix
(Theorem~\ref{thm::intertwining}), that the evolution of
$\mathcal{H}_A(s)$ satisfies the {\em intertwining property}, assuming
$\mathcal{H}(s)$ is non-degenerate, countably dimensional, and twice
differentiable.  The intertwining property can be written
\begin{equation}
  U_A(s)P(0) = P(s)U_A(s)\;,
\end{equation}
where $U_A(s)$ is the unitary operator associated with
$\mathcal{H}_A(s)$, and $P(s)$ is the projection onto the ground state
of $\mathcal{H}(s)$.  Then
\begin{align}
  Q(s)U_A(s)P(0) &= Q(s)U_A(s)P(0)  \\
  &= Q(s)P(s)U_A(s)\\
  &= 0 \;,
\end{align}
since $Q(s)P(s)=0$, so the adiabatic approximation is perfect for
$\mathcal{H}(s)$ if the perturbation term $i/\tau [\dot{P}(s),P(s)]$
is added to it.  Notice further that the perturbation gets arbitrarily
small as $\tau$ grows.

Finally, we also observe that noise that commutes
with the Hamiltonian does not cause any state transitions, because it
has no effect on the eigenstates - in other words, it causes no
coupling between states.  For instance, consider the Hamiltonian on
$N$ particles
\begin{equation}
  \mathcal{H}(s) = M(s) \sum_{j=1}^N \sigma^z_j \;,
\end{equation}
where $M(s)$ is a real scalar function representing
a time-dependent applied magnetic field.  Noise in the
magnetic field $M(s)$ results in a perturbation that commutes with
$\mathcal{H}(s)$, and has no effect on the error of the adiabatic
approximation.

\section{Application to the Spin-1/2 Particle in a Rotating Magnetic Field}
\label{sec::tongEX}

Recently Tong et al. \cite{tong} presented an example of a Hamiltonian
evolution for which the adiabatic approximation performs poorly.  The
Hamiltonian is for a spin-1/2 particle in a rotating magnetic field.
Here we apply AT-Noise to their example.  Their evolving Hamiltonian
is
\begin{align}
  \mathcal{H}(t) &= 
  -\frac{\omega_0}{2}\left( \sigma_x \sin\theta \cos \omega t
  + \sigma_y \sin\theta \sin \omega t + \sigma_z \cos\theta
  \right)
\end{align}
which we represent in the $z$ basis as
\begin{align}
  \mathcal{H}(t)
  &= -\frac{\omega_0}{2}\left(
  \begin{array}{lr}
    \cos \theta & e^{-i\omega t} \sin \theta  \\
    e^{i\omega t} \sin \theta & -\cos \theta
  \end{array} \right) \;.
\end{align}

Suppose $\theta$ is small.  We can think of the time-independent
diagonal component of the Hamiltonian as the intended Hamiltonian, and
the wobbling off-diagonal component as a noise term operating on an
independent timescale.

The eigenstates of $\mathcal{H}(t)$ depend on $t$, but
the eigenvalues do not. So the energy gap is constant
and in fact equal to $\omega_0$.  Thus one might think that the
adiabatic approximation works well, predicting that a particle
starting out in the spin-down state stays in the spin-down state under
this Hamiltonian evolution.

We will see below that if the wobble is at a resonant frequency with
respect to the energy difference between the spin-up and spin-down
states, the wobble induces a complete transition from the spin-down to
spin-up state.  So the adiabatic approximation eventually fails in the
most complete sense possible in this example.  However, we will also
see that the AT-Noise correctly provides an {\em increasing} error
bound with time, because the $s$-derivatives in this example increase
with $\tau$.

We can rewrite the Hamiltonian using $t=s\tau$ as
\begin{equation}
  \label{eqn::tongH}
  \mathcal{H}_{\tau}(s) = -\frac{\omega_0}{2}\left(
  \begin{array}{lr}
    \cos \theta & e^{-i\omega s\tau} \sin \theta  \\
    e^{i\omega s\tau} \sin \theta & -\cos \theta
  \end{array} \right) \;.
\end{equation}
We can also compute the first two derivatives:
\begin{align}
  \frac{d}{ds}\mathcal{H}_{\tau}(s) &= 
  -\frac{\omega \omega_0 \tau \sin \theta}{2}\left(
  \begin{array}{lr}
    0 & -ie^{-i\omega s\tau} \\
    ie^{i\omega s\tau} & 0
  \end{array} \right) \;,
\end{align}
and
\begin{align}
  \frac{d^2}{ds^2}\mathcal{H}_{\tau}(s) &= 
  \frac{\omega^2\omega_0\tau^2 \sin \theta}{2}\left(
  \begin{array}{lr}
    0 & -e^{-i\omega s\tau} \\
    e^{i\omega s\tau} & 0
  \end{array} \right)  \;.
\end{align}
We can compute the norms of these matrices exactly, giving

\begin{align}
  \mathcal{H}_{\tau}^{\dagger}(s)\mathcal{H}_{\tau}(s) 
  &= \frac{\omega_0^2}{4} \left(
  \begin{array}{lr}
    1 & 0 \\
    0 & 1
  \end{array} \right) \;,\\
  \left(\frac{d\mathcal{H}_{\tau}(s)}{ds}\right)^{\dagger}\left(\frac{d\mathcal{H}_{\tau}(s)}{ds}\right) 
     &= -\frac{\omega^2 \omega_0^2\tau^2\sin^2\theta}{4} \left(
     \begin{array}{lr}
       1 & 0 \\
       0 & 1
     \end{array} \right) \;,
\end{align}
and
\begin{align}
  \left(\frac{d^2\mathcal{H}_{\tau}(s)}{ds^2}\right)^{\dagger}\left(\frac{d^2\mathcal{H}_{\tau}(s)}{ds^2}\right)
     &= -\frac{\omega^4 \omega_0^2\tau^4\sin^2\theta}{4} \left(
     \begin{array}{lr}
       1 & 0 \\
       0 & 1
     \end{array} \right) \;.
\end{align}
 Thus we can write
\begin{equation}
  \vnorm{\frac{d^k}{ds^k} \mathcal{H}_{\tau}(s)} 
    \leq \frac{\norm{\omega_0}}{2}
    \left(1+\norm{\omega\sin\theta}\tau+
        \omega^2\norm{\sin\theta}\tau^2\right) \;,
\end{equation}
for all $0\leq s \leq 1$ and for
$k=0,1,2$. Also,
$\gamma_{\tau}(s)=\omega_0$ for any $s,\tau$ \cite{tong}, so we let
$\bar{\gamma}_{noise}=\omega_0$.

Schr\"{o}dinger's equation can also be solved exactly for the
Hamiltonian in Equation~(\ref{eqn::tongH}).  Define
$\bar{\omega}=\sqrt{\omega_0^2+\omega^2+2\omega_0\omega\cos\theta}$.
From Tong et al. \cite{tong} the unitary time evolution operator for
this system is
\begin{align}
  U_{\tau}(t) &= \left( \begin{array}{lr}
    \left( \cos\left(\frac{\bar{\omega}t}{2}\right)+
    i\frac{\omega+\omega_0\cos\theta}{\bar{\omega}}
    \sin\left(\frac{\bar{\omega}t}{2}\right)\right) e^{-i\omega t/2} & 
    i\frac{\omega_0\sin\theta}{\bar{\omega}}
    \sin\left(\frac{\bar{\omega}t}{2}\right) e^{-i\omega t/2}   \\
    i\frac{\omega_0\sin\theta}{\bar{\omega}}
    \sin\left(\frac{\bar{\omega}t}{2}\right) e^{i\omega t/2}     &
    \left( \cos\left(\frac{\bar{\omega}t}{2}\right)-
    i\frac{\omega+\omega_0\cos\theta}{\bar{\omega}}
    \sin\left(\frac{\bar{\omega}t}{2}\right)\right) e^{i\omega t/2}
\end{array} \right)\;.
\end{align} 
Therefore the error operator for the adiabatic approximation is
\begin{align}
  Q(t)U_{\tau}(t)P(0) &= 
  \left( \begin{array}{lr} 0 & 0 \\ 0 & 1 \end{array}\right)
  U_{\tau}(t) \left( \begin{array}{lr} 1 & 0 \\ 0 & 0 \end{array}\right)\\
  &=    
  \left( \begin{array}{lr} 
    0 & 0 \\ 
    i\frac{\omega_0\sin\theta}{\bar{\omega}}
    \sin\left(\frac{\bar{\omega}t}{2}\right) e^{i\omega t/2} & 0
  \end{array}\right) \;,
\end{align} 
so
 \begin{align}
  \vnorm{Q(t)U_{\tau}(t)P(0)} &= \norm{\frac{\omega_0\sin\theta}{\bar{\omega}}
  \sin\left(\frac{\bar{\omega}t}{2}\right)}\;.
\end{align} 
If the perturbation is resonant, then $\omega=-\omega_0\cos\theta$ so
$\bar{\omega} = \norm{\omega_0}\sin\theta$.  Then we have
 \begin{equation}
  \vnorm{Q(t)U_{\tau}(t)P(0)} = \norm{\sin\left(\frac{\omega_0\sin(\theta) t}{2}\right)}\;.
\end{equation} 

As an example, assume that $\theta = 0.001$, $\omega=10$,
$\omega_0=-10$.  Let $\chi(\tau)$ be the error bound defined by the
adiabatic theorem.   Then we can calculate
$\vnorm{P_{\tau}(s)-P(s)}$ exactly to get $\delta_0=\delta_1=0.0005$, and so
we have
\begin{align}
  \chi(\tau) &= 0.00900025 + 0.04 \tau
\end{align}
and
\begin{align}
  \vnorm{Q(s)U_{\tau}(s)P(0)} &= \norm{\sin(0.005 s\tau)} \;.
\end{align} 

\begin{figure}
\vskip 0.1in
\begin{center}
\includegraphics[height=3in]{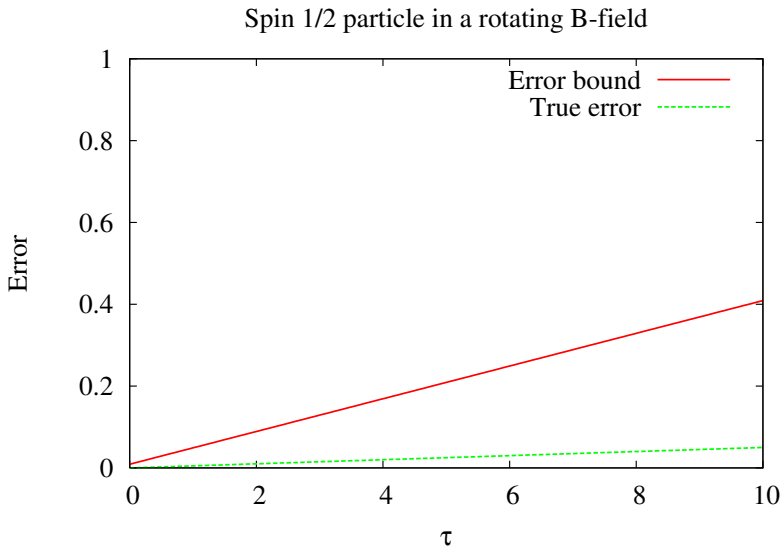}
\end{center}
\caption{A plot of the error bound $\chi(\tau)$ for parameters $\theta
  = 0.001$, $\omega=10$, and $\omega_0=-10$, compared to the true
  error for the Tong et al. \cite{tong} example.  We can see that the
  adiabatic approximation gets worse as $\tau$ gets larger, as
  observed by Tong et al. \cite{tong}.  However, our error bound from
  AT-Noise remains valid.}
\label{fig::chi}
\end{figure}
  
Figure~\ref{fig::chi} illustrates our result.  Not only is the bound
consistent with the true error, it has the same qualitative behavior,
increasing linearly with $\tau$.

\section{Application to a Superconducting Flux Qubit}
\label{sec::fluxqubit}

Next we apply AT-Noise to the superconducting flux qubit of Orlando et
al. \cite{orlando}, proposed for use in adiabatic quantum computation
\cite{lloyd}.  With this qubit, the adiabatic evolution may be as
simple as monotonically varying an applied magnetic field.

\begin{figure}
\begin{center}
\includegraphics[height=3in]{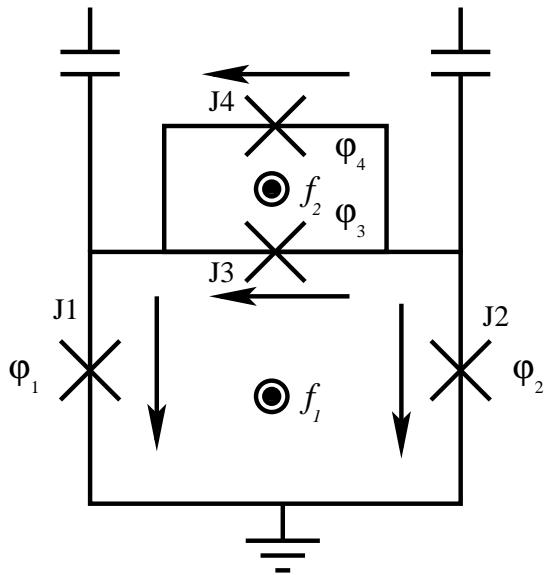}
\end{center}
\caption{(Adapted from Orlando et al. \cite{orlando}) The circuit
schematic for the superconducting flux qubit.  The X's represent
Josephson junctions, and the main qubit loop is formed by junctions
J1, J2, and J3.  Junction J4 allows tuning of the effective properties
of junction J2 through control of the frustration $f_2$.}
\label{fig::circuit}
\end{figure}

Consider the four-junction qubit shown in Figure~\ref{fig::circuit}.
We will follow the analysis of Orlando et al. \cite{orlando}.  The
dynamical variables are the phases $\phi_i$ across the four Josephson
junctions, however flux quantization in each loop gives us two
constraints: $\phi_1-\phi_2+\phi_3=-2\pi f_1$ and $\phi_4-\phi_3=-2\pi
f_2$, where $f_1$ and $f_2$ are the magnetic frustrations in each
loop.  So there are two degrees of freedom, which we define as
$\phi_p=\left( \phi_1+\phi_2\right)/2$ and
$\phi_m=\left(\phi_1-\phi_2\right)/2$, where $f_a = f_2$, and $f_b =
f_1+f_2/2$.  Then the Hamiltonian can be written as
\begin{equation}
  \mathcal{H} = 
  -\frac{\hbar^2}{2M_p}\frac{\partial^2}{\partial \phi_p^2} 
  -\frac{\hbar^2}{2M_m}\frac{\partial^2}{\partial \phi_m^2} +
   U(\phi_p, \phi_m) \;,
\end{equation}
where $M_p$ and $M_m$ are constants,
and the potential $U(\phi_p, \phi_m)$ is defined as
\begin{equation}
  U(\phi_p, \phi_m) = E_J\left[2 + 2\beta - 2\cos(\phi_p)\cos(\phi_m) - 
  2\beta\cos(\pi f_a)\cos(2\pi f_b+2\phi_m)\right] \;,
  \label{eqn::FluxQubitPotential}
\end{equation}
where $E_J$ and $\beta$ are constants.

At $f_1=f_2=1/3$, we have $f_b = 1/2$ and $f_a = 1/3$, and $U(\phi_p,
\phi_m)$ has minima, or {\em wells}, at $\phi_p=0$ and $\phi_m = \pm
\cos^{-1}(\beta/2)$, symmetric about the $\phi_p$ axis.  By varying
$f_1$ and $f_2$, we can tilt the potential so that one well is deeper
than the other, and we can adjust the barrier height.  We can
approximate the Hamiltonian with a two-state system.
A Hamiltonian evolution that begins at the degeneracy
point and varies $f_1$ can can be written:
\begin{align}
  \mathcal{H}(s) &= 
  -t_1\sigma_x
  +  s\epsilon r_1\sigma_z \;,
\end{align}
where $t_1$ and $r_1$ are parameters that can be estimated with the
WKB approximation. For the qubit parameters recommended by Orlando et
al. \cite{orlando}, $r_1 = 4.8E_J$ and $t_1 = 10^{-3}E_J$, where $E_J$
is a constant. A typical value for $E_J$ is $200\; {\rm GHz}\cdot h =
1.256\; {\rm THz}\cdot \hbar$.  We choose $\epsilon = -.0002$, so that
the Hamiltonian changes from proportional to $\sigma_x$ at $s=0$ to
equally-weighted $\sigma_x$ and $\sigma_z$ terms at $s=1$, because
this seems a natural milestone in the evolution to the
$\sigma_z$-dominated final Hamiltonian.

There are a couple of sources of noise
in this qubit.  One source of noise in a superconducting flux qubit is
noise in the critical current of the Josephson
junctions\cite{wellstood}, which decreases as $1/T$ where $T$ is
temperature.  Such noise would result in variations in the weights of
the terms in Equation~(\ref{eqn::FluxQubitPotential}).  Another source
is noise in the magnetic flux bias generated by nearby
current-carrying wires on the chip.  Current carrying wires could be
used for nearby measurement devices or to perform a gate operation on
the qubit with an RF pulse.  Since we have from above a two-state
Hamiltonian parametrized by flux bias, we consider this latter noise.
Orlando et al. \cite{orlando} estimated that a nearby wire of typical
dimensions and carrying 100nA current would cause a difference in
either $f_1$ or $f_2$ of $\Delta f = 10^{-7}$.  Let us assume that
there is there is approximately $0.5\;{\rm nA}$ of noise on the wire
introduced by the current source. Further suppose the
power of the noise scales as inverse
frequency $1/\nu$ from $\nu_{min}=2.5\;{\rm GHz}$ to
$\nu_{max}=3.5\;{\rm GHz}$, so that we include the qubit frequencies
throughout the evolution.

There are clever means of simulating $1/\nu$ noise with discrete
models, such as by summing independent bistable fluctuators
(a.k.a. Random Telegraph Noise) \cite{faoro}.  However, this results
in non-differentiable Hamiltonians so is not appropriate for us.
Instead, suppose we want to write down a formula for a noise source
with $1/\nu$ power spectrum in the range $\nu_{min}$ to
$\nu_{max}$. Let $n$ be an integer, with $n=100$ in the following
example.  Define $\Delta \nu = (\nu_{max}-\nu_{min})/n$ and $\nu_j =
\nu_{min}+j\Delta \nu$ for $j=1...n$.  Then we can define two
independent noise functions, representing variation in the magnetic
frustration in our qubit, as
\begin{align}
  \mathcal{N}_1 (t) = C \sum_{j=1}^n \frac{\cos(2\pi \nu_j t+\xi_{1,j})\Delta \nu}
         {\sqrt{\nu_j}} \;, \\
  \mathcal{N}_2 (t) = C \sum_{j=1}^n \frac{\cos(2\pi \nu_j t+\xi_{2,j})\Delta \nu}
         {\sqrt{\nu_j}} 
\end{align}
where $\xi_{1,j}$ and $\xi_{2,j}$ are phase factors chosen uniformly
at random and $C=10^{-10}\; {\rm MHz^{-1/2}}$, chosen to agree with
the $0.5\;{\rm nA}$ noise.  The Hamiltonian for noise in the magnetic
frustration is
\begin{align}
  \mathcal{H}_{noise}(t) &= \mathcal{N}_1(t)r_1\sigma_z + 
                            N_2(t) (r_2\sigma_z-w\sigma_x) \;,
\end{align}
where $w=2.4E_J$ for the chosen qubit parameters.

Evaluating the functions numerically over an interval much larger than
the longest wavelength reveals the bounds $\norm{\mathcal{N}_i(t)}
\leq 4.9100 \cdot 10^{-10}$, $\norm{\dot{\mathcal{N}}_i(t)} \leq
9.1100 \cdot 10^{-6}\;{\rm MHz}$, and $\norm{\ddot{\mathcal{N}}_i(t)}
\leq 0.1667 \;{\rm MHz}^2$.

Recalling that $t=s\tau$, where $s$ is
unitless, we are ready to compute derivatives and norms
of the whole Hamiltonian:
\begin{align}
  \mathcal{H}_{\tau}(s) &= \mathcal{H}(s) + \mathcal{H}_{noise}(s\tau)\\
  &=  -t_1\sigma_x + sr_1\epsilon\sigma_z 
        + \mathcal{N}_1(t)r_1\sigma_z
	+ \mathcal{N}_2(t)\left(r_2\sigma_z-w\sigma_x\right)  \;,\\
 \vnorm{\dot{\mathcal{H}}_{\tau}(s)} &= 
       \vnorm{ r_1\epsilon\sigma_z
	 + \tau \frac{d}{dt}\mathcal{N}_1(t)r_1\sigma_z
	 + \tau \frac{d}{dt}\mathcal{N}_2(t)(r_2\sigma_z-w\sigma_x) }\;,\\
        &\leq 1206.4 \;{\rm MHz}+\tau \cdot 
	      84.7149 \;{\rm MHz}^2  \;,\\
  \vnorm{ \ddot{\mathcal{H}}_{\tau}(s)} &= \vnorm{ \tau^2
       \frac{d^2}{dt^2}\mathcal{N}_1(t)r_1\sigma_z+ \tau^2
       \frac{d^2}{dt^2}\mathcal{N}_2(t)(r_2\sigma_z-w\sigma_x) }\;, \\
        &\leq \tau^2\;  1.5502\cdot 10^6 \; {\rm MHz}^3 \;.
\end{align}
Observe that since $s$ is unitless, $\mathcal{H}(s)$
and its $s$-derivatives all have units of energy, but since $\hbar=1$,
energy units are inverse time units.

We also need to compute the minimum energy gap.  In this case, it
occurs at $s=0$, and the energy gap is
$\bar{\gamma}_{noise} = 2t_1 = 2513\;{\rm
MHz}$.

Finally, we need to find $\delta_0$ and $\delta_1$.  We compute the
projection operators directly and obtain the bounds $\delta_0 =
1.800\cdot 10^{-6}$ and $\delta_1 = 9.117 \cdot 10^{-7}$.  From
AT-Noise, we have
\begin{equation}
  \vnorm{Q_{\tau}(s)U_{\tau}(s)P(0,\tau)} \leq
	  1.9634 \;\tau + 0.0019 
	+ \frac{0.0148}{\tau} \;,
\end{equation}
where if $\tau$ is supplied in microseconds, the resulting bound is
unitless.

\begin{figure}
\vskip 0.1in
\begin{center}
\includegraphics[height=3in]{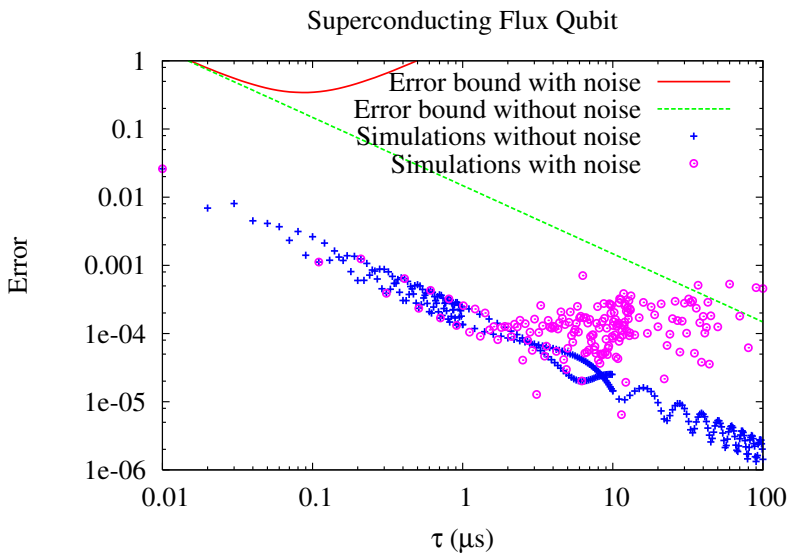}
\end{center}
\caption{A plot of the error bound $\chi(\tau)$ for the
  superconducting flux qubit example.  The crosses represent results
  of numerical simulations of the error. Since the noise commutes with
  the dominant term in the final Hamiltonian, the bound is a
  substantial overestimate.  Nonetheless the qualitative shape between
  the bound curve and the simulation data agree.}
\label{fig::chi2}
\end{figure}

This generates a hyperbolic curve with a vertical asymptote at
$\tau=0$ and a linear asymptote for large $\tau$, shown in
Figure~\ref{fig::chi2}.  Recall this curve represents the norm of the
error operator and its square represents the probability of error in
this system.

To check our results, we would like to compute the error of the
adiabatic approximation numerically.  However efficient numerical
simulation of this system requires some care.  A straightforward
solution to Schr\"{o}dinger's equation
 \begin{equation}
  i\frac{d}{dt}\ket{\psi(t)} = \mathcal{H}_{\tau} (t/\tau)\ket{\psi(t)}
  \label{eq::schrod}
\end{equation} 
in the $\sigma_z$ basis will have rapidly oscillating phases that make
the solutions very unstable and time-consuming to compute.  Instead,
we will rewrite Schr\"{o}dinger's equation for this system in a basis
whose phase rotates with time.

To begin, we choose a time-dependent eigenbasis of
$\mathcal{H}_{\tau}(t/\tau)$ with the property\\
$\braket{\phi_n(t)}{\dot{\phi}_n(t)} = 0$.  In other words,
the Berry phase, also known as geometric
phase, is zero.

\begin{lem}
There is a time dependent eigenbasis $\left\{ \ket{\phi_n(t)}
\right\}$ with the property that
\\$\braket{\phi_n(t)}{\dot{\phi}_n(t)} = 0$ for all $n$ \cite{bozic}.

\begin{proof}
Suppose $\left\{ \ket{\phi_n(t)} \right\}$ does not have this
property.  Then for each $n$, define the variable $\xi_n(t)$ as
follows:
\begin{equation}
  \xi_n(t) = i \int_0^t \braket{\phi_n(w)}{\dot{\phi}_n(w)} dw
\end{equation}
and let 
\begin{equation}
  \ket{\alpha_n(t)} = e^{i\xi_n(t)}\ket{\phi_n(t)} \;.
\end{equation}
Then $\left\{ \ket{\alpha_n(t)} \right\}$ has the desired property.  To
see this, we simply compute it:
\begin{align}
  \braket{\alpha_n(t)}{\dot{\alpha_n}(t)} &=
     \bra{\phi_n(t)}e^{-i\xi_n(t)}
     \left[ i\dot{\xi}_n(t)e^{i\xi_n(t)}\ket{\phi_n(t)}+
            e^{i\xi_n(t)}\ket{\dot{\phi}_n(t)}\right] \\
     &= i \dot{\xi}_n(t) + \braket{\phi_n(t)}{\dot{\phi}_n(t)} \\
     &= i\cdot i\braket{\phi_n(t)}{\dot{\phi}_n(t)} + 
        \braket{\phi_n(t)}{\dot{\phi}_n(t)}\\
     &= 0 \;.
\end{align}
\end{proof}     
\end{lem}

Let us write the solution $\ket{\psi(t)}$ in terms of the basis states
$\ket{\phi_n(t)}$ with energies $E_n(t)$ as follows:
\begin{equation}
  \ket{\psi(t)} = \sum_n c_n(t) e^{-i\int_0^t E_n(w)dw}\ket{\phi_n(t)} \;.
  \label{eq::newbasis}
\end{equation}
Then in the case of two states, assuming the eigenstates are labeled
in increasing order with respect to their eigenvalues, the norm of the
adiabatic error operator is simply $\norm{c_1(t)}$.

Let us substitute the representation in Equation~(\ref{eq::newbasis})
into Schr\"{o}dinger's equation (Equation~(\ref{eq::schrod})), and left
multiply by $\bra{\phi_m(t)}$. After simplification, this yields
\begin{align}
  \dot{c}_m(t) = -\sum_{n} c_n(t) e^{-i\int_0^t \left(E_n(w)-E_m(w)\right)dw}
                \braket{\phi_m(t)}{\dot{\phi}_n(t)}\;.
\end{align}
Evidently the $n=m$ term in the sum is zero in the chosen basis.

Since we have 
\begin{align}
  \mathcal{H}_{\tau}(t/\tau) &= 
     \left(-t_1-\mathcal{N}_2(t)s_2\right) \sigma_x + 
     \left(\frac{t\;r_1\;\epsilon}{\tau} + \mathcal{N}_1(t)r_1 +
           \mathcal{N}_2(t)r_2\right)\sigma_z \;,
\end{align}
if we define
\begin{align}
  a(t) &= -t_1-\mathcal{N}_2(t)s_2 \;, \\ 
  b(t) &=  \frac{t\;r_1\;\epsilon}{\tau} + \mathcal{N}_1(t)r_1 +
  \mathcal{N}_2(t)r_2 \;, \\ 
  \theta(t) &= \cot^{-1}\left(\frac{b(t)}{a(t)}\right) \;,
\end{align}
then we can diagonalize the Hamiltonian easily in terms of $a(t)$ and
$b(t)$.  We choose the cotangent for numerical stability because
$b(0)\approx 0$.
\begin{align}
  E_0(t) &= -\sqrt{a^2(t)+b^2(t)} & 
  E_1(t) &= \sqrt{a^2(t)+b^2(t)}\\
  \ket{\phi_0(t)} &= 
     \left( \begin{array}{c} 
        -\sin \left(\frac{\theta(t)}{2}\right) \\
        \cos \left(\frac{\theta(t)}{2}\right)
      \end{array} \right) &
  \ket{\phi_1(t)} &= 
     \left( \begin{array}{c} 
        \cos \left(\frac{\theta(t)}{2}\right) \\
        \sin \left(\frac{\theta(t)}{2}\right)
      \end{array} \right) \;.
\end{align}
Now, we would like to compute $\braket{\phi_0(t)}{\dot{\phi}_1(t)}$
and $\braket{\phi_1(t)}{\dot{\phi}_0(t)}$.
\begin{align}
  \braket{\phi_0(t)}{\dot{\phi}_1(t)} &=
     \left( \begin{array}{c} 
        -\sin \left(\frac{\theta(t)}{2}\right) \\
        \cos \left(\frac{\theta(t)}{2}\right)
      \end{array} \right) \cdot
     \left( \begin{array}{c} 
       -\frac{1}{2}\sin \left(\frac{\theta(t)}{2}\right) \dot{\theta}(t)\\
       \frac{1}{2}\cos \left(\frac{\theta(t)}{2}\right) \dot{\theta}(t)
      \end{array} \right) \\
     &= \frac{\dot{\theta}(t)}{2}\\
  \braket{\phi_1(t)}{\dot{\phi}_0(t)} &=
     \left( \begin{array}{c} 
        \cos \left(\frac{\theta(t)}{2}\right) \\
        \sin \left(\frac{\theta(t)}{2}\right)
      \end{array} \right) \cdot
     \left( \begin{array}{c} 
       -\frac{1}{2}\cos \left(\frac{\theta(t)}{2}\right) \dot{\theta}(t)\\
       -\frac{1}{2}\sin \left(\frac{\theta(t)}{2}\right) \dot{\theta}(t)
      \end{array} \right) \\
     &= -\frac{\dot{\theta}(t)}{2}
\end{align}  
It remains to compute $\dot{\theta}(t)$, which can be done with
implicit differentiation.
\begin{align}
  \cot (\theta(t)) &= \frac{b(t)}{a(t)}\\
  -\csc^2 (\theta(t)) \dot{\theta}(t) &= \frac{\dot{b}(t)a(t)-b(t)\dot{a}(t)}{a^2(t)} \\
  \dot{\theta}(t) &= \sin^2 (\theta(t)) \frac{\dot{a}(t)b(t)-a(t)\dot{b}(t)}{a^2(t)} \\
  &= \frac{a^2(t)}{a^2(t)+b^2(t)} \frac{\dot{a}(t)b(t)-a(t)\dot{b}(t)}{a^2(t)} \\
  &= \frac{\dot{a}(t)b(t)-a(t)\dot{b}(t)}{a^2(t)+b^2(t)}
\end{align}
Finally, the equations of motion are
\begin{align}
  \dot{c}_0(t) &=  -  c_1(t)e^{-2i\int_0^t \sqrt{a^2(w)+b^2(w)}dw} \frac{\dot{\theta}(t)}{2} \\
  \dot{c}_1(t) &= c_0(t)e^{2i\int_0^t \sqrt{a^2(w)+b^2(w)}dw} \frac{\dot{\theta}(t)}{2}
\end{align}

We provide these equations to a differential equation solver, ode23,
in Matlab.  Care must be taken with the integral in the exponent.  We
need not recompute the integral entirely at each time; rather we cache
the intermediate values of this integral.  Thus at each evaluation we
only integrate on the interval from the last cached time to the
current time.

This method was used to produce the numeric results in
Figure~\ref{fig::chi2}.  The parameters of the system are those
previously described in the example.  There are 571 noiseless data
points and 125 noisy data points.  A new set of random phases was
generated for each noisy point, and each point took up to 5.5 CPU
hours to compute.  The workstation used had dual Xeon 3.06 GHZ
processors with hyperthreading enabled (thus four effective CPUs) and
6GB of RAM, running Red Hat Enterprise 3.

The simulation data in Figure~\ref{fig::chi2} is several orders of
magnitude less than the bound.  The fact that the bound is an
overestimate is not surprising since the noise term commutes with the
dominant term in the final Hamiltonian of the evolution.  However the
qualitative shape between the bound curve and the simulation data is
the same, and the bound does provide a simulation-free guarantee of
error for an interval of $\tau$.

\section{Conclusion}

We provide rigorous bounds for the adiabatic approximation under for
four sources of experimental error: perturbations in the initial
condition, systematic time-dependent perturbations in the Hamiltonian,
coupling to low-energy quantum systems, and decoherent time-dependent
perturbations in the Hamiltonian.

We applied the new results to the spin-1/2 particle in a rotating
magnetic field, which is a standard example for discussing controversy
in the adiabatic theorem \cite{bozic,tong,wuyang}.  We showed that our
theorem makes correct predictions about the error of the adiabatic
approximation as a function of time.

We also applied the new results to the superconducting flux qubit
proposed by Orlando et al. \cite{orlando}, with time-dependent
perturbations in the applied magnetic field.  This qubit has
properties that make it a candidate for quantum adiabatic computation
\cite{lloyd}.  Because our version of the adiabatic theorem does not
have unspecified constants, we are able to make numerical predictions
about this qubit.  We showed that for a particular amount of noise on
superconducting wires near a qubit with ideal physical parameters, we
could guarantee a small error in the adiabatic approximation provided
that the evolution time was set within a particular interval.

\section{Acknowledgments}

The authors would like to thank Stephen Bullock, P. Aaron Lott, Ben
Reichardt, Jocelyn Rodgers, and Eite Tiesinga for helpful comments and
feedback.

{\bf NIST disclaimer.}  Certain commercial products may be identified
in this paper to specify experimental procedures.  Such identification
is not intended to imply recommendation or endorsement by the National
Institute of Standards and Technology.

\appendix
\renewcommand{\thesection}{Appendix A:}
\renewcommand{\theequation}{A-\arabic{equation}}
\setcounter{equation}{0}  
\section{Proof of the Adiabatic Theorem}
\renewcommand{\thesection}{A}
\label{app::at}
Our proof of the adiabatic theorem follows closely those by Avron et
al.  \cite{avron} (later corrections exist \cite{avron2}
\cite{klein}), Reichardt \cite{reichardt}, and Jansen et
al. \cite{ruskai}.  The purpose of revisiting the proof is to have
explicit definitions of constants.

After reviewing some properties of projection operators, we will
introduce the version of Schr\"{o}dinger's equation that will be used
and the assumptions that it requires.  Then we will introduce some
essential lemmas, and finally the proof.

\subsection{Properties of projection operators}

Before embarking on the proof, it will be helpful to review some
properties of projection operators.  First, we will enumerate some
elementary properties.  Then, we will introduce the resolvent
formalism for rewriting projection operators as a contour integral.

Define the commutator $[A,B]$ as $[A,B]=AB-BA$, and $\dot{A} = dA/ds$.
Let $\mathcal{H}(s)$ be some Hamiltonian with countable eigenstates
$\{\ket{\psi_j(s)}\}$ and eigenvalues $\{\lambda_j(s):j\geq 0\}$.  Let
$P(s)$ be the orthogonal projection operator onto the
subspace
\begin{equation}
  \Psi(s) = {\rm Span}\left\{\ket{\psi_m(s)},...,\ket{\psi_n(s)}\right\}\;,
\end{equation}
for some $0 \leq m \leq n$. Thus
\begin{equation}
  P(s)=\sum_{i=0}^k \ket{\psi_i(s)}\bra{\psi_i(s)} \;. 
  \label{eqn::proj}
\end{equation}
Let $Q(s)=I-P(s)$ be the orthogonal complement of $P(s)$.  Then the
following properties hold:
\begin{enumerate}
\item \label{prop::first} $P(s)=P^2(s)$.
\item \label{prop::second} $\dot{P}(s)=\dot{P}(s)P(s)+P(s)\dot{P}(s)$,
  obtained by differentiating both sides of Property
  (\ref{prop::first}).
\item $P(s)\dot{P}(s)P(s)=0$, obtained by multiplying Property
  (\ref{prop::second}) from the left by $P$.
\item $Q(s)P(s)=0$, using the definition of $Q$ and Property
  (\ref{prop::first}).
\item $P^{\dagger}(s)=P(s)$ and $Q^{\dagger}(s)=Q(s)$, where
  $^{\dagger}$ indicates the conjugate transpose.  This is evident
  from Equation (\ref{eqn::proj}).
\item $\vnorm{P(s)} = \vnorm{Q(s)} = 1$.  Recall that the norm of an
  operator $P(s)$ is defined to be the maximum of
  $\vnorm{P(s)\ket{x}}$ for choices of normalized states $\ket{x}$.
  For a projection operator, the maximal choice is a vector in the
  plane of projection, and in that case $P(s)\ket{x}=\ket{x}$.  So
  $\vnorm{P(s)} = 1$.  For $Q(s)$, choose $\ket{x}$ orthogonal to the
  plane of projection.
\item $[\mathcal{H}(s),P(s)]=0$.  To prove this, take some state
 $\ket{\phi}$ and rewrite it as $\ket{\phi} = \sum_{j\geq 0}
 c_j\ket{\psi_j(s)}$.  Then
 \begin{align}
    \mathcal{H}(s)P(s)\ket{\phi} 
    &= \sum_{j\geq 0} c_j\mathcal{H}(s)P(s)\ket{\psi_j(s)} \\
    &= \sum_{j=m}^n c_j\mathcal{H}(s)\ket{\psi_j(s)} \\
    &= \sum_{j=m}^n  c_j\lambda_j(s)\ket{\psi_j(s)}
  \end{align}
  and
  \begin{align}
    P(s)\mathcal{H}(s)\ket{\phi} 
    &= \sum_{j=m}^n P(s)\mathcal{H}(s)c_j\ket{\psi_j(s)} +
                  \sum_{j\not\in [m,n]} P(s)\mathcal{H}(s)c_j\ket{\psi_j(s)}\\
    &= \sum_{j=m}^n c_jP(s)\lambda_j(s)\ket{\psi_j(s)} + 
		 \sum_{j\not\in [m,n]} c_jP(s)\lambda_j(s)\ket{\psi_j(s)}\\
    &= \sum_{j=m}^n c_j\lambda_j(s)\ket{\psi_j(s)} \; .
  \end{align}
\end{enumerate}
 
We will make use of the resolvent formalism to bound the projection
operators.  Define the {\em resolvent} of a Hamiltonian
$\mathcal{H}(s)$ to be
\begin{equation}
  R(z;\mathcal{H}(s)) = (\mathcal{H}(s)-zI)^{-1} \; .
\end{equation}
Suppose we can draw a contour
$\Gamma(s)$ in the complex plane whose
enclosed region includes the eigenvalues corresponding to
$\Psi(s)$ and excludes the rest of the
spectrum of $\mathcal{H}(s)$.  Then we can rewrite the projection
operator $P(s)$ in terms of a line integral of the resolvent $R(s,z) =
(\mathcal{H}(s)-zI)^{-1}$ around this contour:
\begin{equation}
  P(s) = -\frac{1}{2\pi i} \oint_{\Gamma(s)} R(s,z) dz \; .
  \label{eqn::resolvent}
\end{equation}

\subsection{Schr\"{o}dinger's equation}

We rewrite Schr\"{o}dinger's equation in terms of unitary evolution
operators and the scaled time $s$, rather than using state vectors and
real time $t$.  Doing so will introduce the assumption that
$\mathcal{H}(s)$ has a continuous, bounded second derivative.

The usual expression of the time-dependent Schr\"{o}dinger's equation
is
\begin{align}
i\hbar \frac{d\ket{\psi(t)}}{dt} = \mathcal{H}(t)\ket{\psi(t)} \; .
\end{align}
Setting $t = s\tau$, $\ket{\phi(s)}=\ket{\psi(s\tau)}=\ket{\psi(t)}$,
and $\mathcal{H}_{\tau}(s) = \mathcal{H}(s\tau) = \mathcal{H}(t)$, we
can substitute and apply the chain rule for derivatives to get
\begin{align}
\hbar \ket{\dot{\phi}(s)} 
&= -i \tau \mathcal{H}_{\tau}(s)\ket{\phi(s)} \; ,
\end{align}
where the dot indicates the $s$-derivative. It will be
convenient in this proof to choose our units so that $\hbar = 1$.
Also we will assume that all subsequent state vectors, Hamiltonians,
and time evolution operators are functions of the normalized time
parameter $s$, so we can drop the subscript $\tau$ from
$\mathcal{H}_{\tau}$.  Thus we will write
\begin{align}
\ket{\dot{\phi}(s)} &= -i \tau \mathcal{H}(s)\ket{\phi(s)} \; .
\end{align}

Now define $U(s)$ so that for any $\ket{\phi(0)}$, we have
$U(s)\ket{\phi(0)} = \ket{\phi(s)}$ where $\ket{\phi(s)}$ is the
solution to this equation.  Then we proceed as in \cite{sakurai}.
Assume that $\mathcal{H}(s)$ has a continuous bounded derivative; then
$\ket{\phi(s)}$ has a continuous bounded second derivative.  Thus the
remainder for the first-order Taylor
expansion is well-defined.  For some point $s^{*} \in [s, s+\Delta
s]$, we get
\begin{align}
U(s+\Delta s) \ket{\phi(0)} &= \ket{\phi(s+\Delta s)} \\
        &= \ket{\phi(s)} + \ket{\dot{\phi}(s)} \Delta s + 
	\ket{\ddot{\phi}(s^{*})} \frac{\Delta s ^2} {2} \\
        &= \ket{\phi(s)}-i \tau \mathcal{H}(s) \ket{\phi(s)} \Delta s + 
	\mathcal{O}(\Delta s^2)\\
&= U(s)\ket{\phi(0)}-i \tau \mathcal{H}(s) U(s)\ket{\phi(0)} \Delta s 
     + \mathcal{O}(\Delta s^2) \; .
\end{align}
Since this is true for any $\ket{\phi(0)}$ we can write
\begin{align}
  \lim_{\Delta s\rightarrow 0} \frac{U(s+\Delta s)-U(s)}{\Delta s} &=
  -i \tau \mathcal{H}(s) U(s)
\end{align}
or, equivalently,
\begin{align}
  \dot{U}(s) &= -i \tau \mathcal{H}(s) U(s) \; .
  \label{eqn::Schrodinger}
\end{align}
Equation~(\ref{eqn::Schrodinger}) is the form of Schr\"{o}dinger's
equation that we will rely on for the rest of the proof of the
adiabatic theorem.

\subsection{Essential Lemmas}

Recall that the adiabatic approximation states that a system of
Hamiltonian $\mathcal{H}(s)$, initially in some state in $\Psi(0)$,
evolves to approximately some state in $\Psi(s)$ at time $t=s\tau$. To
compute bounds on the error of this approximation, we will identify a
Hamiltonian $\mathcal{H}_A$ that has {\em exactly} this property.
Define
\begin{equation}
  \label{eqn::HA}
  \mathcal{H}_A(s) = \mathcal{H}(s) + \frac{i}{\tau}
  \left[\dot{P}(s),P(s)\right]
\end{equation}
where $P(s)$ is the projection operator onto $\Psi(s)$. Evidently
$\mathcal{H}_A$ is a $1/\tau$ perturbation of $\mathcal{H}$, where
$\tau$ is the scale factor between normalized time and unnormalized
time.  Define $U_A(s)$ to be the unitary evolution operator that is
the solution to Schr\"{o}dinger's equation for $\mathcal{H}_A$, namely
\begin{equation}
  \dot{U}_A(s)=-i\tau \mathcal{H}_A(s)U_A(s) \;.
\end{equation}

The important property of $\mathcal{H}_A(s)$ can be restated as
follows.  If a system is initialized in $\Psi(0)$ at time $s=0$, the
state at time $s$ under evolution by the Hamiltonian
$\mathcal{H}_A(s)$ is entirely contained in $\Psi(s)$.  We can write
this property, known as the {\em intertwining property}, using $P(s)$
and $U_A(s)$ as defined in the previous paragraph.

\begin{thm}[The Intertwining Property]
\label{thm::intertwining}
For all $0\leq s\leq 1$, let $\mathcal{H}(s)$ be Hermitian, twice
differentiable, non-degenerate, and have a countable number of
eigenstates.  Let $U_A(s)$ and $P(s)$ be defined as previously. Then
  \begin{equation}
    U_A(s)P(0) = P(s)U_A(s) \;.
  \end{equation}
\begin{proof}
Noticing that $U_A(s)$ is unitary, we can rewrite the claim as $P(0) =
U_A^{\dagger}(s)P(s)U_A(s)$.  Since $U_A(0) = I$ this is certainly
true for $s = 0$.  So it is sufficient to show that
\begin{align}
  \frac{d}{ds} \left[ U_A^{\dagger}(s)P(s)U_A(s) \right] & = 0 \; .
\end{align}
Applying the product rule for derivatives we get
\begin{align}
  \frac{d}{ds} \left[ U_A^{\dagger}(s)P(s)U_A(s) \right] & = 
      \frac{d}{ds} \left[ U_A^{\dagger}(s)P(s)\right] U_A(s) + 
      U_A^{\dagger}(s)P(s) \dot{U}_A(s) \\
   & = \left( U_A^{\dagger}(s)\dot{P}(s) + \dot{U_A^{\dagger}}(s)P(s) \right) U_A(s)
                 + U_A^{\dagger}(s)P(s) \dot{U}_A(s) \; .
\end{align}
Now observe that $\dot{U_A^{\dagger}} = (\dot{U}_A)^{\dagger}$ since
the derivative of a matrix operator is the derivative of its matrix
entries.  Further, recall that $\dot{U}_A(s) = -i\tau
\mathcal{H}_A(s)U_A(s)$.  So
\begin{align}
  \dot{U_A^{\dagger}}(s) &= (\dot{U}_A(s))^{\dagger} \\
             &= \left(-i\tau \mathcal{H}_A(s)U_A(s)\right)^{\dagger} \\
             &= +i\tau U_A^{\dagger}(s)\mathcal{H}_A^{\dagger}(s) \\
             &= i\tau U_A^{\dagger}(s)\mathcal{H}_A(s) \; ,
\end{align}
since $\mathcal{H}_A$ is Hermitian.  Substituting, we get
\begin{align}
  \frac{d}{ds} \left[ U_A^{\dagger}(s)P(s)U_A(s) \right] =& 
     \left( U_A^{\dagger}(s)\dot{P}(s) + 
     i\tau U_A^{\dagger}(s)\mathcal{H}_A(s) P(s) \right) U_A(s) \notag \\
   & +  U_A^{\dagger}(s)P(s) (-i)\tau \mathcal{H}_A(s)U_A(s) \\
 =& U_A^{\dagger}(s) \left( \dot{P}(s) + i\tau \mathcal{H}_A(s) P(s) - i\tau P(s)\mathcal{H}_A(s) \right) U_A(s)\\
 =& U_A^{\dagger}(s) \left( \dot{P}(s) +  i\tau[\mathcal{H}_A(s), P(s)] \right) U_A(s) \; .
\end{align}
Now we will work on the inner term.  We use the properties that
$[H(s),P(s)]=0$, $P(s)\dot{P}(s)P(s)=0$, $P^2(s)=P(s)$, and
$\dot{P}(s)=\dot{P}(s)P(s)+P(s)\dot{P}(s)$.
\begin{align}
  [\mathcal{H}_A(s), P(s)] =& \mathcal{H}_A(s)P(s) - P(s)\mathcal{H}_A(s) \\
  =& \left(\mathcal{H}(s)+\frac{i}{\tau}
  \left(\dot{P}(s)P(s)-P(s)\dot{P}(s)\right)\right)P(s) \notag \\
  & - P(s)\left(\mathcal{H}(s)+\frac{i}{\tau}
  \left(\dot{P}(s)P(s)-P(s)\dot{P}(s)\right)\right) \\
  =& [\mathcal{H}(s),P(s)]+\frac{i}{\tau}\dot{P}(s)P^2(s)-\frac{i}{\tau}P(s)\dot{P}(s)P(s) \notag \\
  & - \frac{i}{\tau} P(s)\dot{P}(s)P(s) + \frac{i}{\tau} P^2(s)\dot{P}(s) \\
  =& \frac{i}{\tau}\dot{P}(s)P^2(s)  + \frac{i}{\tau} P^2(s)\dot{P}(s) \\
  =& \frac{i}{\tau}\dot{P}(s)P(s)  + \frac{i}{\tau} P(s)\dot{P}(s) \\
  =& \frac{i}{\tau}\dot{P}(s) \; .
\end{align}
We can substitute this into the original expression to get
\begin{align}
  \frac{d}{ds} \left[ U_A^{\dagger}(s)P(s)U_A(s) \right] & = U_A^{\dagger}(s) \left( \dot{P}(s) +  i\tau[\mathcal{H}_A(s), P(s)] \right) U_A(s) \\
& = U_A^{\dagger}(s) \left( \dot{P}(s) - \dot{P}(s) \right) U_A(s) \\
& = 0 \; .
\end{align}
\end{proof}
\end{thm}
Notice that this implies an intertwining property for
the orthogonal complement:
\begin{equation}
  \label{eqn::IP2}
  U_A(s)Q(0) = Q(s)U_A(s) \;.
\end{equation}

In the proof of the adiabatic theorem we will make use of the {\em
twiddle} operation.  For a fixed $s$, let $P$ be a projection operator
onto $\Psi(s)$, and assume the eigenvalues
corresponding to $\Psi(s)$ are separated by a gap from the rest of the
eigenvalues. Define
\begin{equation}
  \tilde{X}(s) = \frac{1}{2\pi i} 
  \oint_{\Gamma(s)}R(s,z) X(s)R(s,z)dz
  \label{eqn::twiddledef}
\end{equation}
where $\Gamma(s)$ is a contour in the
complex plane around the eigenvalues associated with the eigenstates
onto which $P(s)$ projects, whose enclosed region excludes any other
eigenvalues of $\mathcal{H}(s)$. We will need the following property
of the twiddle operation.

\begin{lem}[The Twiddle Lemma]
\label{lem::twiddle}
Assume $\hbar = 1$.  For a fixed $s$, let $P$ be a projection operator
onto $\Psi(s)$, and assume the eigenvalues corresponding to $\Psi(s)$
are separated by a gap from the rest of the eigenvalues.  Define
$Q=I-P$, and let $X$ be a bounded linear operator.  Then
\begin{equation}
  QXP =
  -Q\left( \left[\mathcal{H}_A,\tilde{X}\right] -  
  \frac{i}{\tau}\left[\dot{P},\tilde{X}\right]\right) P
\end{equation}

\begin{proof}

We begin by observing that since $P^2=P$ and $QP=0$,
\begin{align}
  -Q[X,P]P &= -Q(XP - PX)P\\
  &=-QXP \; .
\end{align}
Further, since the identity operator
commutes with everything, $[zI,R(z)XR(z)] = 0$.  Then
\begin{align}
  [\mathcal{H},\tilde{X}] &= 
  \mathcal{H} \left(\frac{1}{2\pi i}\oint_{\Gamma} R(z)XR(z)dz\right)
  - \left(\frac{1}{2\pi i}\oint R(z)XR(z)dz\right) \mathcal{H}\\
  &= \frac{1}{2\pi i}\oint_{\Gamma} \left(\mathcal{H}R(z)XR(z) 
  - R(z)XR(z)\mathcal{H}\right) dz\\
  &= \frac{1}{2\pi i} \oint_{\Gamma} [\mathcal{H}, R(z)XR(z)] dz\\
  &= \frac{1}{2\pi i}\oint_{\Gamma} [\mathcal{H}-zI, R(z)XR(z)] dz
\end{align}
Now we use the fact that $(\mathcal{H}-zI)R=I$, that $X$ does not
depend on $z$, and Equation~(\ref{eqn::resolvent}) to write
\begin{align}
  [\mathcal{H},\tilde{X}] 
  &= \frac{1}{2\pi i}\oint_{\Gamma} \left( XR(z) - R(z)X \right)dz \\
  &= \frac{1}{2\pi i} [X,\oint_{\Gamma} R(z) dz] \\
  &= -[X,P] \; .
\end{align}
Also, using the definition of $\mathcal{H}_A$ in (\ref{eqn::HA}), we
have
\begin{align}
  QXP &= Q[X,P]P\\
  &= -Q[\mathcal{H},\tilde{X}] P\\
  &= -Q\left( \left[\mathcal{H}_A,\tilde{X}\right] - 
     \frac{i}{\tau}\left[\left[\dot{P},P\right],\tilde{X}\right]\right) P \; .
\end{align}

All we need to finish the proof is to show
\begin{equation}
    Q\left[\left[\dot{P},P\right],\tilde{X}\right]P = 
       Q\left[\dot{P},\tilde{X}\right]P \; .
\end{equation}
Using $P\dot{P}P = 0$, $P^2 =P$, and $\dot{P} = \dot{P}P+P\dot{P}$, we
have
\begin{align}
Q[[\dot{P},P]\tilde{X}]P &= Q[\dot{P}P-P\dot{P},\tilde{X}]P\\
&= Q\left(\dot{P}P-P\dot{P}\right)\tilde{X}P - Q\tilde{X}\left(\dot{P}P-P\dot{P}\right)P\\
&= (I-P)\left(\dot{P}P-P\dot{P}\right)\tilde{X}P - (I-P)\tilde{X}\left(\dot{P}P-P\dot{P}\right)P\\
&= \dot{P}P\tilde{X}P - P\dot{P}\tilde{X}P +P\dot{P}\tilde{X}P - \tilde{X}\dot{P}P + P\tilde{X}\dot{P}P\\
&= \dot{P}\tilde{X}P -P\dot{P}\tilde{X}P - \tilde{X}\dot{P}P + P\tilde{X}\dot{P}P\\
&= (I-P)\dot{P}\tilde{X}P - (I-P)\tilde{X}\dot{P}P\\
&= Q\dot{P}\tilde{X}P - Q\tilde{X}\dot{P}P\\
&= Q[\dot{P},\tilde{X}]P \;.
\end{align}
\end{proof}
\end{lem}

\subsection{Completion of the Proof}

Now we are ready to prove the adiabatic theorem.  The operator $Q(s)
U(s) P(0)$ takes the $\Psi(0)$ component
of a state vector, and evolves it for (normalized) time $s$ under
$\mathcal{H}(s)$, then computes the error from the adiabatic
approximation.  The theorem guarantees an upper bound on the norm of
this operator.

\begin{thm}[The Adiabatic Theorem]

Assume for $0\leq s\leq 1$ that $\mathcal{H}(s)$ is twice
differentiable, and let 
\begin{align}
  \vnorm{\dot{\mathcal{H}}(s)} &\leq b_1(s)\;, &
  \vnorm{\ddot{\mathcal{H}}(s)} &\leq b_2(s) \;,
\end{align}
Further assume that $\mathcal{H}(s)$ has a countable number of
eigenstates, with eigenvalues \\$\lambda_0(s)\leq \lambda_1(s)...$,
and that $P(s)$ projects onto the eigenspace associated with the
eigenvalues $\{\lambda_m(s), \lambda_{m+1}(s), ... \lambda_n(s)\}$.
Define
\begin{align}
  w(s) &= \lambda_n(s)-\lambda_m(s) \;,& 
  \gamma(s) &= \left\{ \begin{array}{ll}
    \min \{\lambda_{n+1}(s)-\lambda_n(s), \;
       \lambda_m(s)-\lambda_{m-1}(s)\} & m > 0 \\
       \lambda_{n+1}(s)-\lambda_n(s) & m = 0 
       \end{array}\right. \;, \notag \\
  D(s) &= 1 + \frac{2w(s)}{\pi \gamma(s)} \;,&
  Q(s) &= I - P(s) \;.
\end{align}
Finally, assume $\gamma(s) >0 $ all $s$.  Then we have
\begin{align}
  \vnorm{Q(s)U(s)P(0)} 
 \leq& \frac{8D^2(0)b_1(0)}{\tau\gamma^2(0)}
  + \frac{8D^2(s)b_1(s)}{\tau\gamma^2(s)} \notag \\
  & + \int_{0}^s \frac{8D^2(r)}{\tau\gamma^2(r)}
  \left( \frac{8(1+D(r))b_1^2(r)}{\gamma(r)} + b_2(r)\right) dr \;.
\end{align}

\begin{proof}
By multiplying by the identity and applying
Theorem~\ref{thm::intertwining} (the intertwining property), we can
write:
\begin{align}
  Q(s)U(s)P(0) &= Q(s) U_A(s) U_A^{\dagger}(s) U(s) P(0) \\
   &= U_A(s) Q(0) U_A^{\dagger}(s) U(s) P(0) \;.
  \label{eqn::orig}
\end{align}
Since $\vnorm{U_A(s)}=1$, if $\vnorm{Q(0) U_A^{\dagger}(s) U(s) P(0)}$
is small the magnitude of the error in the adiabatic approximation is
small.  In fact, if we define 
\begin{equation}
  W(s) = U_A^{\dagger}(s) U(s)
\end{equation}
then $W(s)$ satisfies a useful integral equation, and we prove the
adiabatic theorem by bounding $\vnorm{Q(0) W(s) P(0)}$ instead of
working directly on $Q(s)U(s)P(0)$.  To find the integral equation, we
need to compute $\dot{W}(s)$.  Using the product rule for derivatives,
Schr\"{o}dinger's equation, and the definition of $\mathcal{H}_A(s)$
in (\ref{eqn::HA}), we have
\begin{align}
\dot{W}(s) &= U_A^{\dagger}(s)\dot{U}(s) + \dot{U}_A^{\dagger}(s) U(s)\\
           &= -i\tau U_A^{\dagger}(s) \mathcal{H}(s) U(s) + i\tau U_A^{\dagger}(s)
              \mathcal{H}_A(s) U(s) \\
	   &= -U_A^{\dagger}(s) \left[\dot{P}(s),P(s)\right] U(s) \\
	   &= -U_A^{\dagger}(s) \left[\dot{P}(s),P(s)\right] U_A(s) W(s) \;.
	      \label{eqn::dW}
\end{align}
Clearly $W(0) = I$ and so 
\begin{equation}  
  \label{eqn::IntEqn}
  W(s)=I-\int_0^s U_A^{\dagger}(r) [\dot{P}(r),P(r)] U_A(r)W(r)dr
\end{equation}
It will be useful sometimes to refer to the kernel of this integral
equation, so we define 
\begin{equation}
  K(r) = U_A^{\dagger}(r) \left[\dot{P}(r),P(r)\right] U_A(r) \;.
\end{equation}

Now we can use Equation~(\ref{eqn::IntEqn}) to rewrite $\vnorm{Q(0) W(s)
P(0)}$.  Using the fact that $Q(0)P(0) = 0$, we can write
\begin{equation}
  \label{eqn::ATorigInt}
  Q(0)W(s)P(0) = -\int_0^s Q(0) K(r) W(r) P(0) dr \;.
\end{equation}

Our plan is to rewrite the integrand to obtain an expression where all
but one term has a $1/\tau$ factor.  Integration by parts on the
remaining term will ensure all terms have a $1/\tau$ factor.  Then we
can factor out the $1/\tau$ and bound the operators in each term to
yield the Adiabatic Theorem.

To obtain this expression, we need to introduce a $P(0)$ in the middle
of Equation~(\ref{eqn::ATorigInt}) so that we can apply
Lemma~\ref{lem::twiddle}.  To do so, we will use the fact that $Q(0) =
Q(0)^2$ to introduce another $Q(0)$, and then use the fact that $Q(0)
K(r) = K(r) P(0)$.

To show that $Q(0) K(r) = K(r) P(0)$, we use intertwining properties,
the fact that $Q(r)P(r)=0$, and the properties $P^2(r)=P(r)$ and
$P(r)\dot{P}(r)P(r)=0$:
\begin{align}
  Q(0) K(r) &= Q(0) U_A^{\dagger}(r) [\dot{P}(r),P(r)] U_A(r) \\
  &= U_A^{\dagger}(r) Q(r) [\dot{P}(r),P(r)] U_A(r) \\
  &= U_A^{\dagger}(r) \left(Q(r)\dot{P}(r)P(r)-Q(r)P(r)\dot{P}(r)\right) U_A(r) \\
  &= U_A^{\dagger}(r) Q(r)\dot{P}(r)P(r) U_A(r) \\
  &= U_A^{\dagger}(r) \left(\dot{P}(r)P^2(r)-P(r)\dot{P}(r)P(r)\right) U_A(r) \\
  &= U_A^{\dagger}(r) \left[\dot{P}(r),P(r)\right]P(r) U_A(r) \\
  &= U_A^{\dagger}(r) \left[\dot{P}(r),P(r)\right] U_A(r)P(0) \\
  &= K(r) P(0) \;.
\end{align}
Then we can rewrite
\begin{align}
  Q(0)W(s)P(0) &= -\int_0^s Q(0) K(r) W(r) P(0) dr \\
  &= -\int_0^s Q(0)^2 K(r) W(r) P(0) dr \\
  &= -\int_0^s Q(0) K(r) P(0) W(r) P(0) dr \;.
\end{align}
Now we use the definition of $K(r)$, the properties $P^2(r)=P(r)$ and
$Q^2(r)=Q(r)$, and the intertwining property again:
\begin{align}
  Q(0)W(s)P(0) &= -\int_0^s Q(0)^2 U_A^{\dagger}(r) 
                \left[\dot{P}(r),P(r)\right] 
                U_A(r) P(0)^2 W(r) P(0) dr \\
  &= -\int_0^s Q(0) U_A^{\dagger}(r)Q(r) \left[\dot{P}(r),P(r)\right] 
		 P(r)U_A(r) P(0) W(r) P(0) dr \;.
\end{align}
We would like to apply Lemma~\ref{lem::twiddle} with
\begin{align}
  X(r) =& \left[\dot{P}(r),P(r)\right] \;.
\end{align}
In order to apply the lemma, we need to show $X(r)$ is a bounded
linear operator.  Clearly $X(r)$ is linear, and since $P(r)$ has unit
norm then it is sufficient to show that $\vnorm{\dot{P}(r)}$ has a
bound.

\begin{figure}
\vskip 0.1in
\begin{center}
\includegraphics[height=2in]{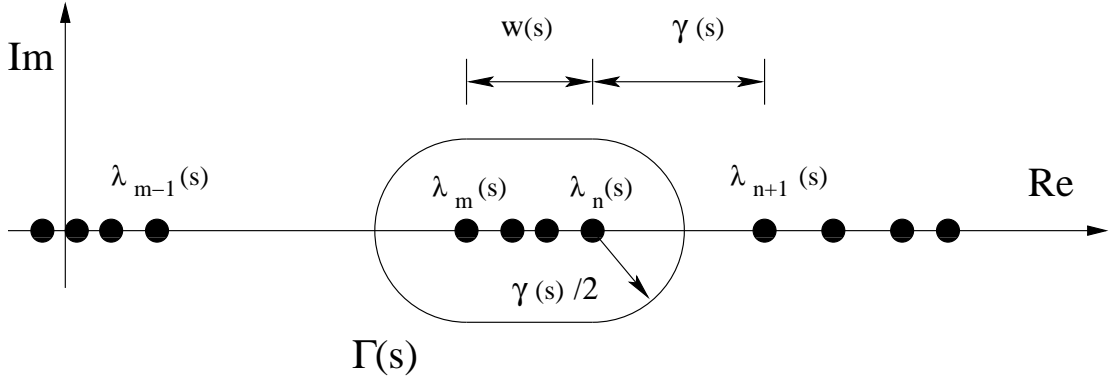}
\end{center}
\caption{Visualization of the resolvent contour $\Gamma(s)$.  The
  eigenvalues of $\mathcal{H}(s)$ (represented with black dots) all
  lie along the real axis since $\mathcal{H}(s)$ is Hermitian.  Notice
  $\Gamma(s)$ lies at least $\gamma(s)/2$ from any eigenvalues.  The
  length of $\Gamma(s)$ is $\pi \gamma(s) + 2w(s) = D(s)\pi\gamma(s)$.
  Observe that $D(s)$ is the ratio of the length of $\Gamma(s)$ to the
  circumference of a circle of radius $\gamma(s)/2$.}
\label{fig::gamma}
\end{figure}

To bound the norm of $\dot{P}(r)$ we will use the resolvent formalism.
We first need to bound the norm of the resolvent $R(r,z)$.  Notice
that if the eigenstates of $\mathcal{H}(r)$ are $\{\ket{\psi_j(r)}:j
\geq 0\}$, then
\begin{equation}
  R(r,z) = \sum_{j\geq 0} \frac{1}{\lambda_j(r)-z}
  \ket{\psi_j(r)}\bra{\psi_j(r)}\;,
\end{equation}
so the norm of $R(r,z)$ equals the inverse of the minimum distance of
$z$ to an eigenvalue of $\mathcal{H}(r)$.  So we need to choose the
contour $\Gamma(r)$ around the eigenvalues of $\Psi(r)$ to maximize
the minimum distance of $\Gamma(r)$ to any eigenvalue.  Also, to
obtain the best bound on the path integral, we will want to minimize
the length of $\Gamma(r)$, given that
maximum minimum distance.  We choose $\Gamma(r)$ consisting of two
semicircles connected by lines, forming a pill-shape.  The semicircles
are centered at $\lambda_m(r)$ and $\lambda_n(r)$, and they have
radius $\gamma(r)/2$.  Figure~\ref{fig::gamma} illustrates this
choice, which bounds the norm of $R(r,z)$ at $2/\gamma(r)$ and the
length at $D(r)\pi\gamma(r)$.

We can check the tightness of this choice by using it to bound the
norm of $P(r)$, which we know is unity.  We have
\begin{align}
  \vnorm{P(s)} &= \vnorm{ -\frac{1}{2\pi i} \oint_{\Gamma(r)} R(r,z) dz}\\
    &\leq \frac{1}{2\pi} D(r) \pi \gamma(r) \frac{2}{\gamma(r)} \\
    &= D(r) \;,
\end{align}
so the approximation is tight for $D(r)=1$.  When $D(r)>1$, it is
complicated by the fact that the closest eigenvalue is not always the
same at different points on $\Gamma(r)$.

The elements of $R(r,z)$ are rational functions of the elements of
$\mathcal{H}(r)$, which are assumed to be differentiable.  So we can
apply the quotient rule for derivatives to determine that $R(r,z)$
is differentiable for $z$ not an eigenvalue of $\mathcal{H}(r)$.

We proceed by differentiating both sides of the equation
\begin{equation}
  R(r,z)(\mathcal{H}(r)-zI) = I
\end{equation}
and multiplying both sides by $R(r,z)$ on the right.  We thus obtain
\begin{align}
  \dot{R}(r,z) &= -R(r,z)\dot{\mathcal{H}}(r)R(r,z) \; .
\end{align}
So by Equation~(\ref{eqn::resolvent})
\begin{align}
  \dot{P}(r) = \frac{1}{2\pi i} 
    \oint_{\Gamma(r)} R(r,z)\dot{\mathcal{H}}(r)R(r,z) dz \;.
  \label{eqn::PDot}
\end{align}

Also, recall that $\vnorm{\dot{\mathcal{H}}(r)} \leq b_1(r)$, so we
can bound the norm of the integral in Equation~(\ref{eqn::PDot}) with
a rectangle approximation.  Using our formula for the length of
$\Gamma(r)$, we get
\begin{align}
  \vnorm{\dot{P}(r)} 
    &\leq \frac{1}{2\pi}D(r)\pi\gamma(r) \frac{4b_1(r)}{\gamma(r)^2}\\
    &= \frac{2D(r)b_1(r)}{\gamma(r)} \;.
\end{align}

Finally, we can bound $\vnorm{X(r)}$.  Using the definition of $X$ we
have
\begin{align}
  \vnorm{X(r)} &= \vnorm{[\dot{P}(r),P(r)]}\\
  &\leq 2\vnorm{\dot{P}(r)}\;\vnorm{P(r)}\\
  &= \frac{4D(r)b_1(r)}{\gamma(r)} \;.
\end{align}

Thus we can apply Lemma~\ref{lem::twiddle}.  We remove the extra
$Q(r)$ and $P(r)$ the same way they were introduced, and use
Schr\"{o}dinger's equation.
\begin{align}
  Q(0)W(s)P(0) 
  =& \int_0^s Q(0) U_A^{\dagger}(r)
  \left( \left[\mathcal{H}_A(r),\tilde{X}(r)\right] - 
         \frac{i}{\tau}\left[\dot{P}(r),\tilde{X}(r)\right]\right)
  U_A(r) P(0) W(r) P(0) dr \notag \\
  =& \int_0^s Q(0) U_A^{\dagger}(r) \mathcal{H}_A(r)\tilde{X}(r) 
      U_A(r) P(0) W(r)P(0)  dr \notag \\
  & - \int_0^s Q(0) U_A^{\dagger}(r) \tilde{X}(r) 
      \mathcal{H}_A(r)U_A(r) P(0)W(r)P(0)  dr \notag \\
  & -\frac{i}{\tau} \int_0^s Q(0) U_A^{\dagger}(r) \left[\dot{P}(r),\tilde{X}(r)\right] 
                             U_A(r)P(0)W(r)P(0) dr\\
  =& \int_0^s Q(0) U_A^{\dagger}(r) \mathcal{H}_A(r)\tilde{X}(r) 
              U_A(r) P(0) W(r)P(0) dr \notag \\
  & -\frac{i}{\tau}\int_0^s Q(0) U_A^{\dagger}(r) \tilde{X}(r) 
	                    \dot{U}_A(r) P(0)W(r)P(0) dr \notag \\
  & -\frac{i}{\tau} \int_0^s Q(0) U_A^{\dagger}(r) \left[\dot{P}(r),\tilde{X}(r)\right] 
                             U_A(r)P(0)W(r)P(0) dr\;.
			     \label{eqn::preIntByParts}
\end{align}

Evidently the last two integrals have a $1/\tau$ factor, and we only
need to work on the first integral.  We will integrate it by parts,
using
\begin{align}
A(r) &= \tilde{X}(r)U_A(r)P(0)W(r) \\
dA &= \tilde{X}(r)U_A(r)P(0)\dot{W}(r)dr + 
      \left(\dot{\tilde{X}}(r)U_A(r)+\tilde{X}(r)\dot{U}_A(r)\right)P(0) W(r)dr \\
B(r) &= U_A^{\dagger}(r) \\
dB &= i\tau U_A^{\dagger}(r)\mathcal{H}_A(r)dr \;.
\end{align}
Applying the integration by parts to $\int dB A$ yields
\begin{align}
\int_{0}^s U_A^{\dagger}(r) \mathcal{H}_A(r)\tilde{X}(r)U_A(r) P(0)W(r) dr =& 
\left. -\frac{i}{\tau} U_A^{\dagger}(r)\tilde{X}(r)U_A(r)P(0)W(r) \right|_{r=0}^s \notag \\
& +\frac{i}{\tau} \int_{0}^s U_A^{\dagger}(r) \tilde{X}(r)U_A(r)P(0)\dot{W}(r) dr \notag \\
& +\frac{i}{\tau} \int_{0}^s U_A^{\dagger}(r)\dot{\tilde{X}}(r)U_A(r)P(0)W(r) dr \notag \\
& +\frac{i}{\tau} \int_{0}^s U_A^{\dagger}(r)\tilde{X}(r)\dot{U}_A(r)P(0)W(r) dr \; .
\end{align}
When we substitute, we see that the last integral cancels with the
second integral in Equation~(\ref{eqn::preIntByParts}), so we obtain
\begin{align}
Q(0)W(s)P(0) =& 
\left. -\frac{i}{\tau} Q(0) U_A^{\dagger}(r)\tilde{X}(r)U_A(r)
     P(0)W(r)P(0) \right|_{r=0}^s \notag \\
& + \frac{i}{\tau} Q(0) 
  \int_{0}^s U_A^{\dagger}(r) \tilde{X}(r)U_A(r)P(0)\dot{W}(r)P(0) dr \notag \\
& +\frac{i}{\tau} Q(0) \int_{0}^s U_A^{\dagger}(r)
    \left(\dot{\tilde{X}}(r)-[\dot{P}(r),\tilde{X}(r)] \right) U_A(r)P(0)W(r)P(0) dr \; .
  \label{eqn::finalInt}
\end{align}

To finish the proof, we need to bound each of the three terms on the
right.  We will do this by applying the triangle inequality to all the
operators in Equation~(\ref{eqn::finalInt}).  Unitary operators and
projection operators have unit norm, and we have bounded $\dot{P}(r)$
already, so it remains to bound the norms of $\tilde{X}(r)$,
$\dot{\tilde{X}}(r)$, and $\dot{W}(r)$.  As dependencies we will also
need to find the norms of $\ddot{P}(r)$ and $\dot{X}(r)$.

\begin{enumerate}
\item $\vnorm{\ddot{P}(r)}$: 
  To bound $\ddot{P}(r)$ we need to compute $\ddot{R}(r,z)$.  Using
  the product rule for derivatives,
  \begin{align}
    -\ddot{R}(r,z) =& \frac{d}{dr} R(r,z)\dot{\mathcal{H}}(r)R(r,z) \\
    =& \frac{d}{dr} \left( R(r,z)\dot{\mathcal{H}}(r)\right)R(r,z) 
       + R(r,z)\dot{\mathcal{H}}(r)\dot{R}(r,z) \\
    =& \left(\dot{R}(r,z)\dot{\mathcal{H}}(r)+R(r,z)\ddot{\mathcal{H}}(r) \right) 
       R(r,z) + R(r,z)\dot{\mathcal{H}}(r)\dot{R}(r,z) \\
    =& \dot{R}(r,z)\dot{\mathcal{H}}(r)R(r,z)+ R(r,z)\ddot{\mathcal{H}}(r)R(r,z) +
       R(r,z)\dot{\mathcal{H}}(r)\dot{R}(r,z) \;.
  \end{align}
  Since $\vnorm{\dot{R}(r)} \leq
  \vnorm{R(r)}^2\vnorm{\dot{\mathcal{H}}(r)} \leq
  4b_1(r)/\gamma(r)^2$, we have
  \begin{align}
    \vnorm{\ddot{R}(r,z)} 
     &\leq \frac{16b_1(r)^2}{\gamma(r)^3}+\frac{4b_2(r)}{\gamma(r)^2} \\
     &= \frac{4}{\gamma(r)^2}\left(\frac{4b_1(r)^2}{\gamma(r)}+b_2(r)\right) \;.
  \end{align}
  So, following the reasoning used to bound $\vnorm{\dot{P}(r)}$,
  \begin{align}
    \vnorm{\ddot{P}(r)} 
    &= \vnorm{ -\frac{1}{2\pi i} \oint_{\Gamma(r)} \ddot{R}(r,z)dz} \\
    &\leq \frac{1}{2\pi} \pi\gamma(r) D(r) \frac{4}{\gamma(r)^2}
    \left(\frac{4b_1(r)^2}{\gamma(r)}+b_2(r)\right)\\
     &= \frac{2D(r)}{\gamma(r)}\left(\frac{4b_1(r)^2}{\gamma(r)}+b_2(r)\right) \;.
  \end{align}
\item $\vnorm{\tilde{X}(r)}$:  
  By Equation (\ref{eqn::twiddledef}), we have
  \begin{align}
    \vnorm{\tilde{X}(r)} & \leq \frac{1}{2\pi} \oint_{\Gamma(r)}
    \vnorm{R(r,z)}\;\vnorm{X(r)}\;\vnorm{R(r,z)} dz\\ &\leq
    \frac{1}{2\pi}\pi\gamma(r) D(r) \frac{2}{\gamma(r)}
    \frac{4D(r)b_1(r)}{\gamma(r)} \frac{2}{\gamma(r)}\\ &=
    \frac{8D(r)^2b_1(r)}{\gamma(r)^2} \;.
  \end{align}
\item $\vnorm{\dot{X}}$: 
  Notice $\dot{X}(r) = \left[\ddot{P}(r),P(r)\right]$, so
  \begin{align}
    \vnorm{\dot{X}(r)} &= \vnorm{[\ddot{P}(r),P(r)]}\\
    &\leq 2\vnorm{\ddot{P}(r)}\;\vnorm{P(r)}\\
    &= \frac{4D(r)}{\gamma(r)}\left(\frac{4b_1(r)^2}{\gamma(r)}+b_2(r)\right)\;.
  \end{align}
\item $\vnorm{\dot{\tilde{X}}(r)}$: 
  We have
  \begin{align}
    \vnorm{\dot{\tilde{X}}(r)} =& \frac{1}{2\pi}\vnorm{\frac{d}{dr} 
                                  \oint_{\Gamma(r)} R(r,z)X(r)R(r,z) dz}\\
    =& \frac{1}{2\pi}\vnorm{\oint_{\Gamma(r)} \frac{d}{dr} R(r,z)X(r)R(r,z) dz }\\
    =& \frac{1}{2\pi}\vnorm{\oint_{\Gamma(r)} R(r,z)X(r)\dot{R}(r,z) + 
               \left(\dot{R}(r,z)X(r)+R(r,z)\dot{X}(r)\right)R(r,z) dz} \notag \\
    =& \frac{1}{2\pi}\left|\left|\oint_{\Gamma(r)} -R(r,z)X(r)R(r,z)\dot{\mathcal{H}}(r)R(r,z) - 
          R(r,z)\dot{\mathcal{H}}(r)R(r,z)X(r)R(r,z) \right.\right. \notag \\
    & \left.\left. + R(r,z)\dot{X}(r))R(r,z) dz \right|\right| \; .
  \end{align}
  Now since $\vnorm{R(r,z)}\leq 2/\gamma(r)$, we get
  \begin{align}
    \vnorm{\dot{\tilde{X}}(r)} 
    &\leq \frac{1}{2\pi}\pi\gamma(r) D(r) 
          \left(\frac{16b_1(r)}{\gamma(r)^3}\vnorm{X(r)} + 
          \frac{4}{\gamma(r)^2}\vnorm{\dot{X}(r)}\right)\\
    &= \frac{8D(r)^2}{\gamma(r)^2}\left(\frac{8b_1(r)^2}{\gamma(r)}
	  +b_2(r)\right) \;.
  \end{align}
\item $\vnorm{\dot{W}(r)}$: 
  Recall from Equation~(\ref{eqn::dW}) that
  \begin{align}
    \dot{W}(r) =& -U_A^{\dagger}(r) \left[\dot{P}(r),P(r)\right]U(r) W(r) \;.
  \end{align}
  We know that $\vnorm{W(r)}=\vnorm{U_A^{\dagger}(r)}=\vnorm{U(r)} =
  1$, and remember that $X(r) = \left[\dot{P}(r),P(r)\right]$.  So we
  can apply the triangle inequality to get
  \begin{align}
    \vnorm{\dot{W}(r)}&\leq \vnorm{X(r)} \\
    &\leq \frac{4D(r)b_1(r)}{\gamma(r)} \;.
  \end{align}
\end{enumerate}

The resulting bounds are
\begin{align}
  \vnorm{\dot{P}(r)} &\leq \frac{2D(r)b_1(r)}{\gamma(r)} \;,
    & \vnorm{\tilde{X}(r)} &\leq \frac{8D(r)^2b_1(r)}{\gamma(r)^2} \;, \\
  \vnorm{\dot{W}(r)} &\leq \frac{4D(r)b_1(r)}{\gamma(r)} \;,
  & \vnorm{\dot{\tilde{X}}(r)}
  &\leq \frac{8D(r)^2}{\gamma(r)^2}\left(\frac{8b_1(r)^2}{\gamma(r)}+b_2(r)\right) \;.
\end{align}

Now let us apply these bounds to Equation~(\ref{eqn::finalInt}) by
taking the norm of both sides:
\begin{small} \begin{align}
\vnorm{Q(0) W(s) P(0)} &\leq 
\frac{1}{\tau} \vnorm{ \left. Q(0)U_A^{\dagger}(r)\tilde{X}(r)U_A(r)P(0)W(r)P(0) 
                       \right|_{r=0}^s } \notag \\
& + \frac{1}{\tau} \vnorm{ Q(0) \int_{0}^sU_A^{\dagger}(r)\tilde{X}(r)U_A(r)P(0) 
                          \dot{W}(r)P(0) dr } \notag \\
& + \frac{1}{\tau} \vnorm{ Q(0) \int_{0}^sU_A^{\dagger}(r)(\dot{\tilde{X}}(r)+
                                [\dot{P}(r),\tilde{X}(r)])U_A(r)P(0) W(r) P(0) dr } \;.
\end{align} \end{small}

We can further simplify this by noting that the norm of each integral
is less than the integral of the norm of its integrand.  Further, we
use the triangle inequality and the fact that the norm of unitary
operators and projection operators are unity:
\begin{align}
  \vnorm{Q(0) W(s) P(0)} \leq& 
 \frac{1}{\tau} \left( 
      \vnorm{\tilde{X}(0)} + \vnorm{\tilde{X}(s)} \right. \notag \\
      &\left. +\int_{0}^s
      \vnorm{\tilde{X}(r)}\vnorm{\dot{W}(r)}
      + \vnorm{\dot{\tilde{X}}(r)}+\vnorm{[\dot{P}(r),\tilde{X}(r)]} dr \right)\\
 \leq& \frac{8D^2(0)b_1(0)}{\tau\gamma^2(0)}
  + \frac{8D^2(s)b_1(s)}{\tau\gamma^2(s)} \notag \\
  & + \int_{0}^s \frac{8D^2(r)}{\tau\gamma^2(r)}
  \left( \frac{8(1+D(r))b_1^2(r)}{\gamma(r)} + b_2(r)\right) dr \;.
\end{align}

Finally, from Equation~(\ref{eqn::orig}), we get
\begin{align}
  \vnorm{Q(s)U(s)P(0)} 
  &\leq \vnorm{Q(0) W(s) P(0)} \\
 \leq& \frac{8D^2(0)b_1(0)}{\tau\gamma^2(0)}
  + \frac{8D^2(s)b_1(s)}{\tau\gamma^2(s)} \notag \\
  & + \int_{0}^s \frac{8D^2(r)}{\tau\gamma^2(r)}
  \left( \frac{8(1+D(r))b_1^2(r)}{\gamma(r)} + b_2(r)\right) dr \;.
\end{align}
We also know that $\vnorm{Q(s)U(s)P(0)}\leq 1$ by the triangle
inequality.
\end{proof}
\end{thm}

\bibliography{aqc}

\end{document}